\documentclass[10pt,journal,compsoc]{IEEEtran}
\usepackage[T1]{fontenc}

\AtBeginDocument{%
  \providecommand\BibTeX{{%
    \normalfont B\kern-0.5em{\scshape i\kern-0.25em b}\kern-0.8em\TeX}}}
\usepackage{cite}
\usepackage{booktabs}  
\usepackage{subcaption} 
\usepackage{amsmath}
\usepackage{chngpage}
\usepackage{multirow}
\usepackage{amssymb}
\usepackage{mdframed}
\usepackage{mathtools}
\usepackage{proof}
\usepackage{bussproofs}
\usepackage{lipsum} 
\usepackage{floatflt,amsmath,amssymb}
\usepackage[ligature, inference]{semantic}
\usepackage{paralist}
\usepackage{cuted}
\usepackage{array}
\usepackage{tikz}
\usepackage{colortbl}
\usepackage{flushend}
\usepackage{scalerel}
\usepackage[]{xcolor}
\usepackage[most]{tcolorbox}
\usepackage{multirow, makecell}
\usepackage{graphicx}
\usepackage{url}
\usepackage{hyperref}
\usepackage{tikz}
\newcommand{\tabincell}[2]{\begin{tabular}{@{}#1@{}}#2\end{tabular}}
\newcommand*\circled[1]{\tikz[baseline=(char.base)]{
            \node[shape=circle,draw,inner sep=0pt] (char) {#1};}}
\newtcbox{\dbox}[1][]{
  math upper,
  baseline=0.4\baselineskip,
  equal height group=dashedbox,
  nobeforeafter,
  colback=white,
  boxrule=0pt,
  enhanced jigsaw,
  boxsep=0pt,
  top=0.5pt,
  bottom=0.5pt,
  left=1pt,
  right=1pt,
  borderline horizontal={0.5pt}{0pt}{dashed},
  borderline vertical={0.5pt}{0pt}{dashed},
  drop lifted shadow,
  #1
}
\newtcbox{\yzxbox}[1][]{
  math upper,
  baseline=0.4\baselineskip,
  equal height group=dashedbox,
  nobeforeafter,
  colback=white,
  boxrule=0pt,
  enhanced jigsaw,
  boxsep=0pt,
  top=0.5pt,
  bottom=0.5pt,
  left=1pt,
  right=1pt,
  borderline horizontal={0.5pt}{0pt}{},
  borderline vertical={0.5pt}{0pt}{},
  drop lifted shadow,
  #1
}

\usepackage{ifthen}
\newboolean{showcomments}
\setboolean{showcomments}{true}
\ifthenelse{\boolean{showcomments}}
 { \newcommand{\mynote}[2]{
      \fbox{\bfseries\sffamily\scriptsize#1}
        {\small$\blacktriangleright$\textsf{\emph{#2}}$\blacktriangleleft$}}}
        { \newcommand{\mynote}[2]{}}

\newcommand\mathbox[1]{\mathord{\ThisStyle{%
  \fboxsep3\LMpt\relax\kern1\LMpt\fbox{$\SavedStyle#1$}\kern1\LMpt}}}

\renewcommand{\theenumii}{\theenumi.\arabic{enumii}.}
\definecolor{primary}{RGB}{117,112,179}
\definecolor{secondary}{RGB}{27,158,119}
\definecolor{tertiary}{RGB}{217,95,2}

\definecolor{orange}{RGB}{255,127,0}
\definecolor{grey}{RGB}{135,135,135}

\lstdefinelanguage{diff}{
  language=java,
  basicstyle=\ttfamily\scriptsize,
  sensitive=true,
  morecomment=[f][\color{gray}][0]{diff},
  morecomment=[f][\color{gray}][0]{index},
  morecomment=[f][\color{blue}][0]{@@},
  morecomment=[f][\color{magenta}][0]{***},
  morecomment=[f][\color{violet}][0]{!},
  morecomment=[f][\color{red!60!black}][0]{-},
  morecomment=[f][\color{green!60!black}][0]{+},
  morecomment=[f][\color{magenta}][0]{---},
  morecomment=[f][\color{magenta}][0]{+++},
  morecomment=[f][\color{gray}][0]{Binary},
  morecomment=[f][\color{gray}][0]{Only},
  morecomment=[f][\color{gray}][0]{old},
  morecomment=[f][\color{gray}][0]{new},
  morecomment=[f][\color{gray}][0]{rename},
  morecomment=[f][\color{gray}][0]{similarity},
  morecomment=[f][\color{gray}][0]{deleted},
  morecomment=[f][\color{magenta}][0]{***************},
  morecomment=[f][\color{red!60!black}][0]<,
  morecomment=[f][\color{green!60!black}][0]>,
  morecomment=[f][\color{blue}][0]{0},
  morecomment=[f][\color{blue}][0]{1},
  morecomment=[f][\color{blue}][0]{2},
  morecomment=[f][\color{blue}][0]{3},
  morecomment=[f][\color{blue}][0]{4},
  morecomment=[f][\color{blue}][0]{5},
  morecomment=[f][\color{blue}][0]{6},
  morecomment=[f][\color{blue}][0]{7},
  morecomment=[f][\color{blue}][0]{8},
  morecomment=[f][\color{blue}][0]{9},
}[comments]

\begin{document}

\title{Learning the Relation between Code Features and Code Transforms with Structured Prediction}

\author{Zhongxing~Yu,
        Matias~Martinez,
        Zimin~Chen,
        Tegawendé~F.~Bissyandé
        and Martin~Monperrus
\IEEEcompsocitemizethanks{\IEEEcompsocthanksitem Z. Yu is with Shandong University.
E-mail: zhongxing.yu@sdu.edu.cn 
\IEEEcompsocthanksitem 
M. Martinez is with  Universitat Politècnica de Catalunya.
E-mail: matias.martinez@upc.edu
\IEEEcompsocthanksitem Z. Chen and M. Monperrus are with KTH Royal Institute of Technology. 
E-mail: zimin@kth.se and monperrus@kth.se
\IEEEcompsocthanksitem T. F. Bissyandé is with University of Luxembourg.
E-mail: tegawende.bissyande@uni.lu
}
\thanks{}}

\markboth{To Appear in IEEE TRANSACTIONS ON SOFTWARE ENGINEERING}%
{Yu \MakeLowercase{\textit{et al.}}}

\IEEEtitleabstractindextext{%
\begin{abstract}
To effectively guide the exploration of the code transform space for automated code evolution techniques, we present in this paper the first approach for structurally predicting code transforms at the level of AST nodes using conditional random fields (CRFs). Our approach first learns offline a probabilistic model that captures how certain code transforms are applied to certain AST nodes, and then uses the learned model to predict transforms for arbitrary new, unseen code snippets. {Our approach involves a novel representation of both programs and code transforms. Specifically, we introduce the formal framework for defining the so-called AST-level code transforms and we demonstrate how the CRF model can be accordingly designed, learned, and used for prediction}. We instantiate our approach in the context of repair transform prediction for Java programs. Our instantiation contains a set of carefully designed code features, deals with the training data imbalance issue, and comprises transform constraints that are specific to code. We conduct a large-scale experimental evaluation based on a dataset of bug fixing commits from real-world Java projects. The results show that when the popular evaluation metric \emph{top-3} is used, our approach predicts the code transforms with an accuracy varying from 41\% to 53\% depending on the transforms. Our model outperforms two baselines based on history probability and neural machine translation (NMT), suggesting the importance of considering code structure in achieving good prediction accuracy. In addition, a proof-of-concept synthesizer is implemented to concretize some repair transforms to get the final patches. The evaluation of the synthesizer on the Defects4j benchmark confirms the usefulness of the predicted AST-level repair transforms in producing high-quality patches.
\end{abstract}

}

\maketitle

\section{Introduction}

\noindent
During the life-cycle of a computer program, its source code evolves under a sequence of transforms. 
Those code transforms are not random: they capture the evolution of the program (e.g., new features and bug fixes) and they also reflect the semantic constraints of the programming language (each version must compile, so not all transforms are acceptable). In other words, there is a probability distribution over the code transform space. In this paper, we address the problem of capturing the probability distribution of code transforms that underlies software evolution. 

Code transform prediction is an important problem. 
It has implications in research areas dealing with automated code evolution, including, for example, program synthesis \cite{gulwani2017}, program repair \cite{Prophet}, super-optimization \cite{Denali} and refactoring \cite{refactoring}, and can be viewed as the foundation to achieve effective search. For instance, in program repair, the probability distribution of code transforms can not only enable a more focused exploration of the search space, but can also result in better patches \cite{Prophet}.

The problem of computing the probability distribution of code transforms given a program is an unsolved problem. Yet, it is being indirectly investigated in the specific application domain of automated program repair \cite{Le2016HDRepair,feedbackgeneration,sequencer}. For instance, the Prophet repair system \cite{Prophet} analyses a set of past commits extracted from version control systems in order to compute the likelihood of a given patch. 
Indeed, the fundamental issue behind automated code evolution is a representation problem: we need to identify the proper representation for both the program and code transforms. If the representation is too fine-grain (e.g., at the token level), one would need a tremendous amount of data or memory to capture the probability distribution. If the representation is too coarse-grain (e.g., at the statement level), the relation between code and code transforms becomes vague and unactionable, making it useless for driving automated code evolution.

In this paper, we propose an approach to learn the relation between code and code transforms. This approach is novel and effective. Its novelty lies in the representation of both programs and code transforms. 

\textbf{Representing programs.} Programs are represented with a combination of abstract syntax trees (AST) and rich carefully engineered and powerful features. This representation has two advantages: (1) the learning algorithm has access to the full program information (all tokens), as well as to the AST tree structure and the AST node types; 
(2) the learning algorithm does not have to extract the probability from scratch, it can leverage the human knowledge encoded in the features to better and faster identify the signal from the noise in the learning data. 

\textbf{Representing code transforms.} To make the code transforms to be precise enough to be automatically applied, we define AST-level code transforms, i.e., code transforms that are attached to specific AST nodes of the code. The AST-level code transforms are defined as follows. An 'edit' is a basic tree edit operation performed on nodes of the abstract syntax tree of the program before change. A `diff' is the complete set of edits done in an atomic code change, as captured by a commit in a version control system. The AST-level code transform is an abstract view over a single edit or a group of several conceptually related edits. {Our intuition is that this abstraction, compared with concrete AST edits, brings more accuracy and scalability for the arguably hard learning task of predicting code changes. We give the conceptual formal framework for defining this novel type of code transform.}

\textbf{Learning algorithm.} The learning machinery is provided by structured prediction \cite{bakir2007predicting}, in particular conditional random fields (CRFs) \cite{CRF1}. Structured prediction is a branch of machine learning that is well suitable for tree-based data such as abstract syntax tress. CRFs recently have been successfully used on programs, including, for example, automatic deobfuscation \cite{property-predict-popl} and automatic renaming \cite{ASTRepresentPLDI}. While there exist neutral network based approaches (\emph{e.g.}, graph neural network \cite{graphnn}) that can also be used for structured prediction, CRF model has the following advantages: (1) CRF model is fully explainable and the chain of reasoning can be looked at through the associated graph; (2) CRF model exploits domain knowledge encoded into feature functions, thus it enables to capture human insight about independence, causality and other subtle relationships in the underlying graph structure; (3) CRF model has significantly fewer parameters which then can be estimated reliably from less data. In our case, the learned CRF model establishes a probabilistic model that captures how certain code transforms are applied to certain AST nodes, and can be used to predict code transforms for arbitrary new, unseen code snippets. {While there exist works that build a model about code transforms by treating code as token sequences \cite{tufano2019empirical,pu2016sk_p,deepdelta} or works that extract useful edit information from few highly similar edit instances \cite{MengPLDI,meng2013lase,rolim2017learning,koyuncu2020fixminer}, our established CRF model accounts the inherent structural dependencies between transforms applied to different code elements and is able to extract and assemble useful information from diverse (i.e., not that similar) example edit instances. The resultant probabilistic model thus is powerful in predicting our newly defined AST-level code transforms.}

We instantiate our approach in the context of Java programs and code transforms used for repairing programs (hereafter referred to as repair transform for brevity). Our prototype system takes as input a set of past bug-fixing commits and produces a probabilistic model that can be used to predict the repair transforms needed for repairing a bug. {We evaluate our approach on the \emph{September 2015/GitHub} dataset offered by Boa \cite{boa,megadiff} (we hereafter call it Boa dataset for brevity)} which contains 4,590,679 bug fixing commits, and measure to what extent our approach correctly predicts the repair transforms to be applied on the program version before the commit. We perform two series of experiments, one on ``single-transform'' diffs and one on "multiple-transform" diffs, and compare our model performance with that of two baselines which respectively use history probability and neural machine translation (NMT) to predict repair transforms. Intuitively speaking, ``single-transform'' diffs and "multiple-transform" diffs respectively refer to the case where one and multiple repair transforms are needed to change the buggy code into the correct code. For "single-transform" diffs, when the popular evaluation metric \emph{top-3} is used \cite{deepdelta, popl16martin}, our overall best performance model achieves 53\% accuracy, and the history probability baseline and NMT baseline accuracies are 31\% and 47\% respectively. To our knowledge, the "multiple-transform" prediction problem is novel and has not been studied so far. It is arguably a harder prediction problem because the prediction space is orders of magnitude large. For "multiple-transform" diffs, when the popular evaluation metric \emph{top-3} is used \cite{deepdelta, popl16martin}, our best performance model achieves 41\% accuracy, and the history probability baseline and NMT baseline accuracies are 0\% and 36\% respectively. We also systematically investigate the impact of configuration parameter, training data size, and feature functions on model performance. Overall, the results show that our established model achieves good prediction accuracy and consistently performs better than the two baselines. Compared to the baselines, our model takes into account the inherent code structure and thus achieves a better prediction performance.

In addition, to further illustrate the usefulness of the predicted AST-level repair transforms in producing the final patch, we implement a proof-of-concept synthesizer which concretizes 8 of the 16 considered repair transforms to generate patches. We evaluate the synthesizer on the widely used Defects4j (v1.2) benchmark, and compare the repair results with those obtained by 5 state-of-the-art deep learning (DL) based automatic program repair (APR) techniques, including CODIT \cite{CODIT}, CoCoNut \cite{CoCoNuT}, DLFix \cite{DLFix}, CURE \cite{jiang2021cure}, and Recoder \cite{Recoder}. The results show that by applying the proof-of-concept synthesizer to concretize the top 1 repair transform prediction by our model, the synthesizer correctly repairs 16 bugs, and a significant number of the repaired bugs are not repaired by the compared DL-based APR techniques. In particular, compared with CODIT, CoCoNut, DLFix, CURE, and Recoder, the proof-of-concept synthesizer repairs 13, 10, 4, 6, and 2 unique bugs respectively. In addition, as the predicted repair transforms are attached to specific AST nodes, the patch synthesis can proceed in a highly focused way. The synthesized patches thus will suffer less from the overfitting issue \cite{YuEmSE}, and our evaluation confirms this. Overall, the evaluation results confirm the usefulness of the predicted AST-level repair transforms in producing high-quality patches. 

To sum up, our contributions are:

\begin{itemize}
    \item A conceptual framework for defining AST-level code transform and a novel approach to predict AST-level code transform based on structured prediction. Both of them are completely novel to the best of our knowledge.
    
    \item An instantiation of the approach for repair transform prediction for Java programs, which contains a set of carefully designed code features and deals with the training data imbalance and transform constraint issues that arise for repair transform prediction problems.
    The prototype implementation is publicly available at \url{https://github.com/zhongxingyu/Seer.git}.
   
    \item A large-scale and systematic experimental evaluation of the approach on the Boa dataset \cite{boa, megadiff}. The results show that our overall best performance model achieves good accuracy of code transform prediction, and consistently performs better than two baselines based on history probability and neural machine translation (NMT).
    
   \item An implementation of a proof-of-concept synthesizer which concretizes some repair transforms to get the final patches, and an evaluation of the synthesizer on the Defects4j benchmark. The evaluation results demonstrate that high-quality patches can be obtained on top of the predicted AST-level repair transforms.
\end{itemize}

The remainder of this paper is structured as follows. We first use a working example to illustrate our approach in Section 2. Section 3 gives a necessary background about abstract syntax tree (AST) and conditional random fields (CRFs). Section 4 introduces the approach to structurally predict AST-level code transform using CRFs, followed by Section 5 which gives an instantiation of the approach in the context of repair transform for Java programs. Section 6 presents a detailed evaluation of the approach, including the proof-of-concept synthesizer. Finally, we discuss some closely related work in Section 7 and conclude the paper in Section 8. 

\section{Overview} 
\begin{figure*}
\centering
 \includegraphics[scale=0.26]{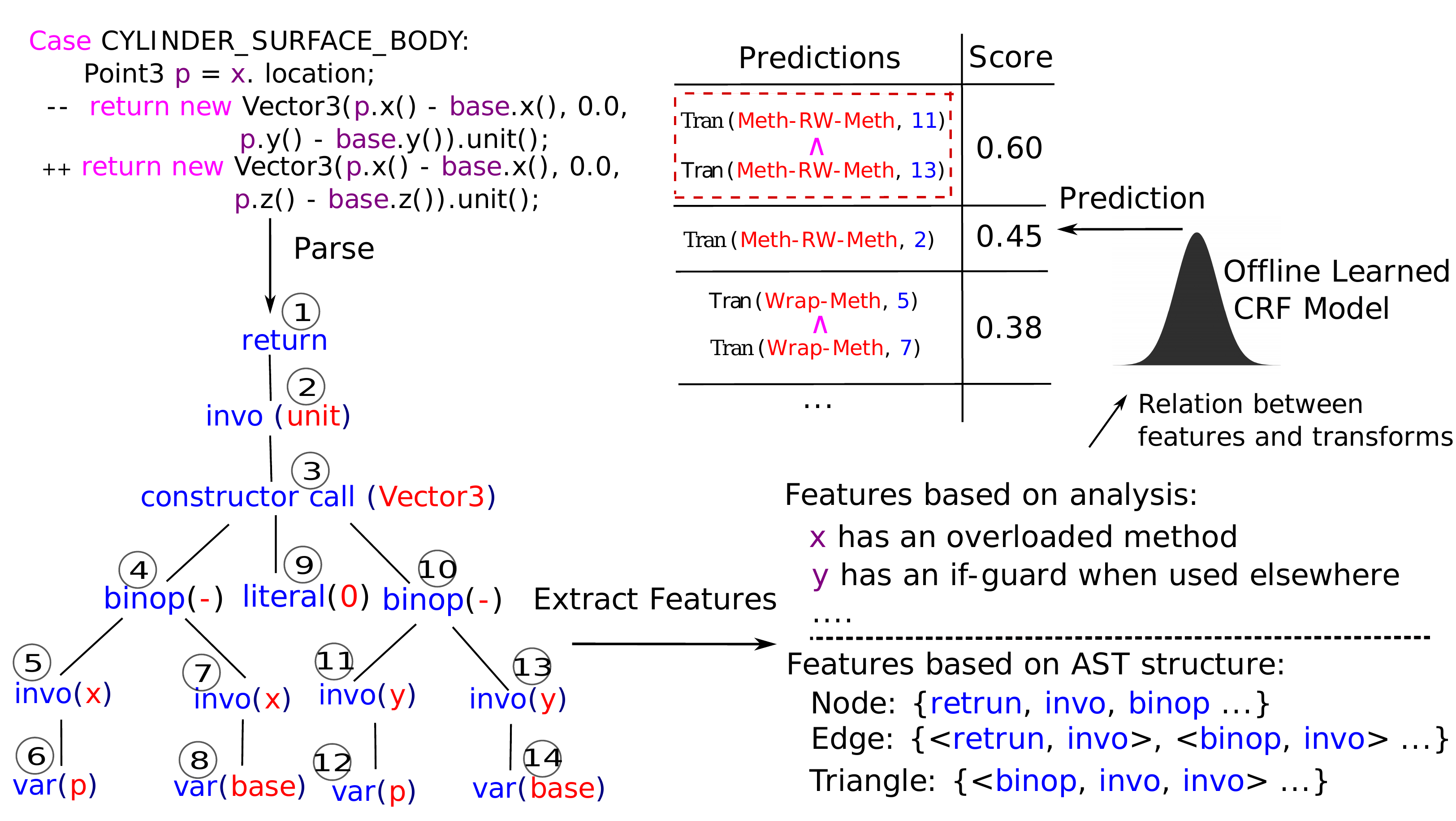}
 \vspace{0.6\baselineskip}   
 \caption{
  Overview of our code transform prediction approach. 
}
 \label{fig:overview}
\end{figure*}
\noindent
In this section, using a working example, we provide an informal overview of our approach for predicting code transforms on AST nodes. \autoref{fig:overview} gives a graphical overview. It shows the diff of Git commit ecc184b in project Jmist\footnote{\url{https://github.com/bwkimmel/jmist/commit/ecc184bc08ee08159cdd79045c2ed0c4245ba59c}}. The problem involves two wrong invocations to method $\mathit{y()}$ that should be replaced by invocations to method $\mathit{z()}$. The code transform behind this diff is a replacement of a method invocation by another one, which is called a "{\tt Meth-RW-Meth}" repair transform in this paper. In the diff of \autoref{fig:overview}, there are two instances of this code transform and they are applied to two different AST nodes. 

The table at the top right hand side of \autoref{fig:overview} shows the predictions of our model. {Each prediction contains a set of predicates $\tt{Tran}$(\emph{T}, \emph{N}), which evaluates to $\tt{true}$ when there exists code transform $\emph{T}$ on AST node \emph{N} (see definition 4.5 for details).} The prediction with the highest score 0.6 is composed of two "{\tt Meth-RW-Meth}" code transforms on AST nodes identified with indexes 11 and 13. Since it is the actual repair changes to be made, it means that for this example, our approach successfully predicts the code transforms. Note that the prediction involves the locations to apply the code transforms: "{\tt Meth-RW-Meth}" code transform points to the two AST nodes corresponding to the invocation of $\mathit{y()}$. We now outline how our approach achieves this.

\vspace{1mm}
\noindent
\textbf{Feature Extraction.} Given the buggy code snippet, our approach first parses it to construct an AST and then extracts the following two types of features:

\begin{itemize}
\item The first type of feature is based on the characteristics of program elements. For instance, they can be whether the invocation $\mathit{x()}$ has overloaded methods and whether the invocation $\mathit{y()}$ is wrapped with an if-check when called in other statements. These features are engineered, and are related with code idioms, semantics not directly captured by the AST (e.g. method overloaded), and common usage. 

\item The second type of feature is based on the abstract syntax tree. All AST nodes are represented with special vertices, edges, and triangles that are used for structured prediction. For the specific example in \autoref{fig:overview}, an excerpt of this representation is shown on the bottom right hand side. 
\end{itemize}

\vspace{1mm}
\noindent
\textbf{Code Transforms.} To effectively guide the code evolution process, the code transforms are defined in terms of their changes on the AST structure. The AST nodes of the buggy code snippet are then annotated with labels that indicate the presence of a code transform. 

\vspace{1mm}
\noindent
\textbf{Offline Model Learning.} After extracting the features for all samples in a training dataset of patches and annotating AST nodes of the buggy code snippets with code transforms, our approach feeds them to a probabilistic model. More specifically, we learn a conditional random field from the data. The learning process makes use of the two types of features mentioned above, and establishes the relation between the features and the code transforms on the AST nodes. In particular, the CRF model allows us to conveniently make the establishment on top of the extracted two types of features using respectively observation-based and indicator-based feature functions.
Put simply, observation-based feature functions facilitate us to establish the correlation between characteristics of program elements and the code transforms on program elements. For example, as overloaded methods are frequently mixed by developers, the correlation between "invocation $\mathit{y()}$ has overloaded methods" and "$\mathit{y()}$ should subject {\tt Meth-RW-Meth} repair transform" should be high. Indicator-based feature functions instead facilitate us to establish the correlation between program structure and the code transforms on program elements through historical information on a large dataset. For instance, in case many instances of the node label edge <invo, var> (like edges \circled{\scriptsize 11} $\rightarrow$ \circled{\scriptsize 12} and \circled{\scriptsize 13} $\rightarrow$ \circled{\scriptsize 14} in \autoref{fig:overview}) in the dataset are associated with repair transform pair <{\tt Meth-RW-Meth}, VOID> where VOID denotes that no repair transform is needed, then the correlation between <invo, var> and <{\tt Meth-RW-Meth}, VOID> should be relatively high. To more accurately establish the correlation, we have designed various type related, usage related, and syntax related code element characteristics (Section 5.2.1) and considered structure-transform relation for AST vertices, edges, and triangles on top of an adequately large, representative dataset (Section 5.2.2). 

The model is learned offline once and the learned weights for the corresponding feature functions reflect the strength of the correlation, and then the model can be used to do predictions for arbitrary, unseen buggy code snippets. 

\vspace{1mm}
\noindent
\textbf{Prediction.} Finally, using the extracted features for the new, unseen buggy code snippet, the already learned model assigns likely code transforms to AST nodes. Each assignment comes with a score representing the probability of the transform. For the buggy code snippet shown in \autoref{fig:overview}, the top-3 predictions by our trained model are shown in the table at the right hand side. The most likely prediction with the score of 0.6 says there is a need to apply two transforms "{\tt Meth-RW-Meth}" (replace one invocation by another one) to AST nodes with indexes 11 and 13 respectively. This prediction is indeed correct. To repair this bug, we exactly need those two repair transforms suggested by the most likely prediction. {With this prediction, we can then employ customized effective synthesis algorithm like \cite{gvero2015synthesizing, perelman2012type} to come up with the actual invocations to replace the two invocations $\mathit{y()}$ at AST nodes 11 and 13.}

\vspace{1mm}
\noindent
\textbf{Key Points.} We now emphasize the key points of our approach. First, our model performs structured prediction and hence predicts the code transforms for all AST nodes of the buggy code snippet given as input. We have a special code transform called EMPTY which means no code transform, and we call the other transforms actual code transforms. The EMPTY predictions are not shown in \autoref{fig:overview}, yet they are indeed outputs of the model. Second, our model does prediction of transforms on specific AST nodes, i.e., the prediction is in a targeted manner. For instance, there are three method invocations involved in the buggy code snippet (i.e., $unit()$, $x()$, $y()$), the most likely prediction attaches the repair transforms to the actual buggy invocation (i.e., the two calls to $y()$). Third, our model can effectively deal with the case when there need multiple actual repair transforms to different AST nodes. Those joint transforms are learned at training time and given as outputs at predication time, as shown in \autoref{fig:overview}. 

\section{Preliminaries} 
\noindent
Before describing our approach in detail, we first provide the necessary background. Our prediction is at the level of AST nodes and we start by formally defining the AST. 

\vspace{1.0mm}
\noindent 
\textbf{Definition 3.1.} \textbf{(Abstract Syntax Tree)}. The abstract syntax tree (AST) for a code snippet is a tuple $<N, T, r, \delta, L, l, V, v>$ where \emph{N} is a set of nonterminal nodes, \emph{T} is a set of terminal nodes, $\emph{r} \in \emph{N}$ is the root node, $\delta : \emph{N} \to (\emph{N} \cup \emph{T})^{\ast} $ is a function that maps a nonterminal node to its children nodes, \emph{L} is a set of node labels, $ l: (\emph{N} \cup \emph{T}) \to \emph{L} $ is a function that maps a node to its label, \emph{V} is a set of node values, and $ v: (\emph{N} \cup \emph{T}) \to (\emph{L} \cup \epsilon) $ is a function that maps a node to its value (can be empty).

Labels of nodes correspond to the names of their production rules in the grammar, i.e., they encode the structure. Values of the nodes correspond to the actual tokens in the code. For instance, for the AST node identified with \circled{10} in \autoref{fig:overview}, its label and value are "{\tt BinaryOperator}" and "-" respectively. 

We use conditional random fields (CRFs) \cite{CRF1} to do the learning. Before describing CRFs in detail, we first give the definition of clique and maximal clique, which are key for understanding CRFs.  

\vspace{1.0mm}
\noindent 
\textbf{Definition 3.2.} \textbf{(Clique and Maximal Clique)}. For an undirected graph \emph{G} = (\emph{V}, \emph{E}), a clique \emph{C} is a set \emph{X} of vertices of \emph{G} such that every two distinct vertices are adjacent. A maximal clique is a clique that cannot be extended by including one more adjacent vertex. 

Imagine an undirected triangle graph, then there are three 1-vertex cliques (the vertices, a special kind of clique), three 2-vertex cliques (the edges), and one 3-vertex clique (the triangle), and the 3-vertex clique is the maximal clique. 

We now give the definition of CRFs. {\emph{Probabilistic graphical models (PGM) use a graph-based representation as the basis for compactly encoding a complex distribution over a high-dimensional space} \cite{PGM}, and CRFs belong to this formalism for expressing the dependence structure of entities. Traditionally, graphical
models have been used to explicitly model the joint probability distribution \emph{p}(\textbf{y}, \textbf{x}) over our observed knowledge about the entities (i.e., \textbf{x}) and the predicted assignment of attributes for the entities (i.e., \textbf{y}). This kind of model is called \emph{generative} model. The limitation of the generative model is that it requires modeling the marginal probability \emph{p}(\textbf{x}), which can be difficult and computationally expensive as the dimensionality of \textbf{x} can be very large and the features may have complex dependencies. An alternative solution to this problem is a \emph{discriminative} model, which models the conditional distribution \emph{p}(\textbf{y}|\textbf{x}) directly and is the approach taken by CRFs. A CRF is a conditional distribution \emph{p}(\textbf{y}|\textbf{x})
with an associated graphical structure, which combines the ability of graphical models to compactly model multivariate outputs \textbf{y} with that of discriminative classification to perform prediction using a large number of input features \textbf{x}.} CRFs have been successfully used in many areas, including, for example, information retrieval \cite{CRFapplication2}, natural language processing \cite{CRFnlp}, bioinformatics \cite{CRFbiomedical}, and computer vision \cite{CRFapplication1}.
The formal definition of CRFs is as follows \cite{CRF1}. 

\vspace{1.0mm}
\noindent  
\textbf{Definition 3.3.} \textbf{(Conditional Random Fields \cite{CRF1})} Let \textbf{X} = $\{\emph{X}_{1}, ... ,  \emph{X}_{N}\}$ and \textbf{Y} = $\{\emph{Y}_{1}, ... , \emph{Y}_{N}\}$ be two sets of random variables, \textbf{x} and
\textbf{y} be realizations of \textbf{X} and \textbf{Y} respectively, \emph{G} = (\emph{V}, \emph{E}) be an undirected graph over \textbf{Y} such that \textbf{Y} is indexed by the vertices of \emph{G}, and \emph{C} be the set of all cliques in \emph{G}. Then (\textbf{X}, \textbf{Y}) is a \emph{conditional random field} if for any value \textbf{x} of \textbf{X} (i.e., conditioned on \textbf{X}), the distribution \emph{p}(\textbf{y}|\textbf{x}) factorizes according to \emph{G} and is represented as:
\begin{equation}
\begin{gathered}
\emph{p}(\textbf{y} | \textbf{x}) = \frac{1}{Z(\textbf{x})} \prod_{{c} \in {C}} \psi_\emph{c}   (\textbf{y}_{c}, \textbf{x})
\end{gathered}
\end{equation}
\normalsize

\noindent where $\emph{Z}(\textbf{x}) = \sum_{\textbf{y} \in {\Omega}_{\textbf{x}}} \prod_{{c} \in {C}} \psi_\emph{c}(\textbf{y}_{c}, \textbf{x})$ is a normalization factor. Here ${\Omega}_{\textbf{x}}$ denotes the set of possible assignments of \textbf{y} for \textbf{x}. {The undirected graph encodes the qualitative aspects of the distribution and edges embody direct dependencies. In addition, note that the factorization in CRFs implicitly assumes a node \textbf{X} and edges (\textbf{X}, $\emph{Y}_{i}$) in the undirected graph.}

Each $\psi_\emph{c}(\textbf{y}_{c}, \textbf{x})$ is a local function that defines the quantitative aspect of the distribution. $\psi_\emph{c}(\textbf{y}_{c}, \textbf{x})$ depends on the whole \textbf{X} but only on a subset $\textbf{Y}_{c} \subseteq \textbf{Y}$ which belong to the clique \emph{c}, and its non-negative scalar value can be deemed as a measure of how compatible the values $\textbf{y}_{c}$ are with each other. Like Hidden Markov Models (HMMs), each local function has the special log-linear form over a prespecified set of feature functions \{\emph{{f}$_k$}\}$_{k=1}^K$:
\begin{equation}
\begin{gathered}
\psi_\emph{c}(\textbf{y}_{c}, \textbf{x}) = \text{exp} \bigg\{\sum_{k=1}^K \lambda_{k}f_{k}(\textbf{y}_{c}, \textbf{x})\bigg\}
\end{gathered}
\end{equation}
\normalsize

Each feature function \emph{{f}$_k$}: $\textbf{Y}_{c} \times \textbf{X}  \rightarrow \mathbb{R}$ is used to score assignments of subset variables $\textbf{Y}_{c}$, and $\lambda_{k}$ is the model parameter to learn which represents the weight for feature function \emph{{f}$_k$}. Typically, the same set of feature functions with the same parameters is used for every clique in the graph. 

\vspace{1.0mm}
\noindent 
\textbf{Example.} Consider the Part-of-Speech (POS) Tagging problem where the goal is to label a sentence with tags like ADJECTIVE, NOUN, etc. For instance, given the sentence "Bob drank coffee at Starbucks", the labeling will be "NOUN VERB NOUN PREPOSITION NOUN". For this problem, the input random variables are the words in a sentence and the output variables are the corresponding tags. The underlying graph could be a liner-chain that connects each $\emph{Y}_{i}$ with $\emph{Y}_{i+1}$, encoding the fact that the labels to different words are dependent on each other. {For liner-chain CRF model, each feature function basically is a function that takes in as input: (1) a sentence \textbf{s}; (2) the position \emph{i} of a certain word in \textbf{s}; (3) the tag {\textbf{t}}$_i$ of the current word; (4) the tag {\textbf{t}}$_{i-1}$ of the previous word. For instance, one possible feature function could be "\emph{{f}$_k$}(\textbf{s}, \emph{i}, {\textbf{t}}$_i$, {\textbf{t}}$_{i-1}$) = 1 if {\textbf{t}}$_{i-1}$ = ADJECTIVE and {\textbf{t}}$_i$ = NOUN, 0 otherwise". If the weight $\lambda_{k}$ associated with this feature is large and positive, the underlying meaning is that adjectives tend to be followed by nouns. The weight for each feature function could be learned during training and then be employed during prediction. Given a sentence \textbf{s}, its tagging \textbf{t} can be scored by adding up the weighted features over all words in \textbf{s}:
\begin{equation}
\begin{gathered}
\emph{score}(\textbf{t} | \textbf{s}) = \sum_{j=1}^m \sum_{i=1}^n \lambda_{j}f_{j} (\textbf{s}, \emph{i}, \textbf{{t}$_i$}, \textbf{{t}$_{i-1}$})
\nonumber
\end{gathered}
\end{equation}
\normalsize
\noindent where the first sum iterates over each feature function \emph{j}, and the inner sum iterates over each position \emph{i} of the sentence. Considering all possible taggings, the scores can be transformed into probabilities  \emph{p}(\textbf{t}|\textbf{s}) by exponentiating and normalizing:
\begin{equation}
\begin{gathered}
\emph{p}(\textbf{t} | \textbf{s}) = \frac{\text{exp} \bigg\{\emph{score}(\textbf{t} | \textbf{s})\bigg\}}{\sum_{\textbf{t}^{'}}\text{exp} \bigg\{\emph{score}(\textbf{t}^{'} | \textbf{s})\bigg\}} 
\nonumber
\end{gathered}
\end{equation}}

\section{Approach for Structured Prediction of Code Transform} 
\noindent
In this section, we introduce our approach for structured prediction of AST-level code transform using conditional random field (CRFs). We first introduce a novel CRF specifically designed for code transform prediction on AST nodes, and then discuss how the approach can be achieved in a step-by-step manner.

\subsection{CRF for Transform Prediction on AST Nodes} 
\noindent
To effectively guide the code transform process, our aim is to predict the needed transforms at the AST node level. We thus first define the following two random fields. 

\vspace{1.0mm}
\noindent 
\textbf{Definition 4.1.} \textbf{(Random field of AST nodes and  transforms)}. 
Let \emph{P} = \{1, 2, 3, ..., \emph{Q}\} be the set of AST nodes where the integer is a unique identifier to denote the nodes traversed in pre-order. We associate a random field \textbf{N} = $\{\emph{N}_{1}, ... , \emph{N}_{Q}\}$ over node symbol (a node symbol is a unique combination of node label and node value) in each position \emph{p} in the AST and another random field \textbf{T} = $\{\emph{T}_{1}, ... , \emph{T}_{Q}\}$ over the code transform applied to each position \emph{p} in the AST.

\begin{figure*}
\centering
 \includegraphics[scale=0.28]{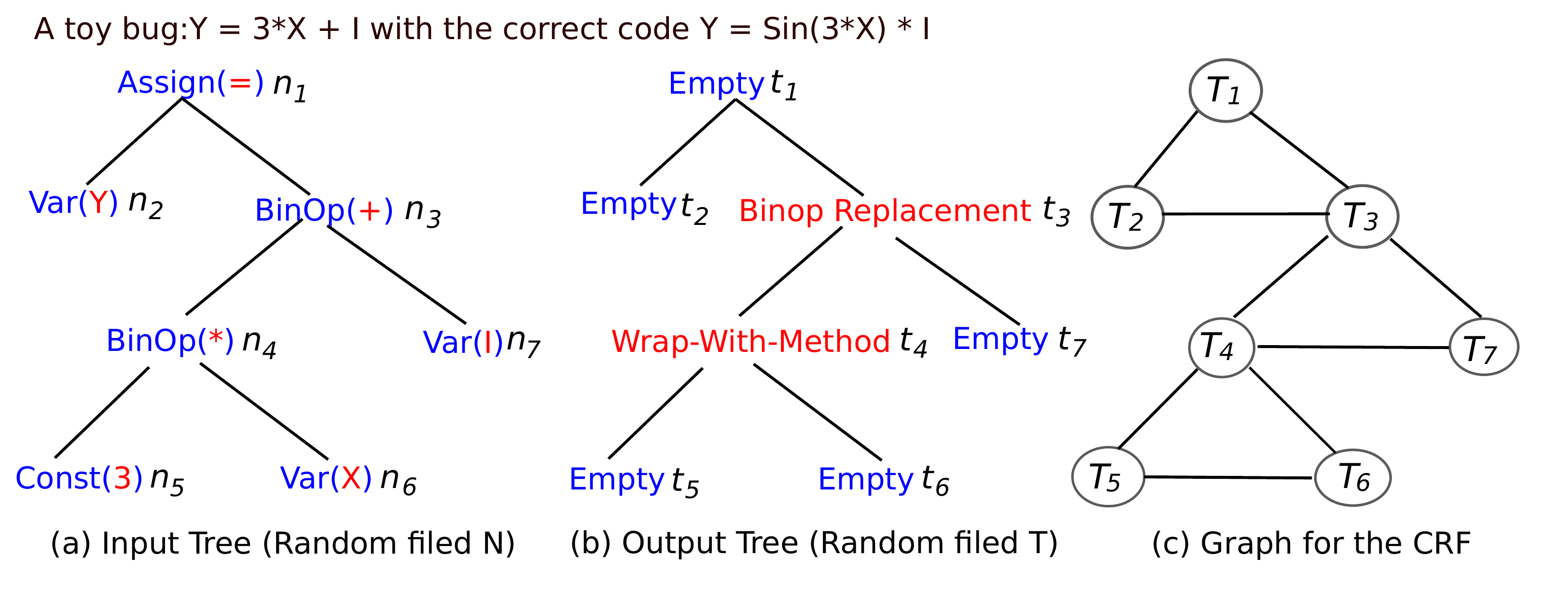}
 \caption{
  An example of input tree, output tree, and CRF graph.
}
 \vspace{-0.8\baselineskip}
 \label{fig:example}
\end{figure*}

The realizations of \textbf{N} (denoted by \textbf{n}) will be the actual nodes for a specific input AST, and the realizations of \textbf{T} (denoted by \textbf{t}) will be the actual transforms applied to nodes of the specific input AST. \autoref{fig:example} (a) and \autoref{fig:example}(b) give an example of the realizations of the two random fields where the transforms are applied to repair the bug. To repair the toy bug in \autoref{fig:example}, we need to replace the binary operator `+' with the binary operator `*' (called "{\tt Binop Replacement}" transform) and wrap expression \emph{3*X} with a method call $\mathit{Sin}$ (called "{\tt Wrap-With-Method}" transform). According to definition 4.1, for realizations of random field \textbf{T}, these two repair transforms are attached to the AST nodes corresponding to binary operator `+' and expression \emph{3*X} in the AST respectively.

Given the two random fields \textbf{N} and \textbf{T}, we now discuss the choice of the undirected graph over random variables in \textbf{T}, i.e., the choice of CRFs for  transform prediction purpose. 
In CRFs, the more complex the graph, the more kinds of feature functions which relate all variables in a clique can be defined, which in turn would lead to a larger class of conditional probability distributions. However, note meanwhile a complex graph will make the exact inference algorithms become intractable \cite{CRF2}.
For the transform prediction problem, we define the undirected graph in the following way:

\vspace{1.0mm}
\noindent  
\textbf{Definition 4.2.} \textbf{(Graph for random field T)} The undirected graph for random field \textbf{T} = $\{\emph{T}_{1}, ... , \emph{T}_{Q}\}$ is \emph{G} = (\emph{V}, \emph{E}) such that (1) \emph{V} = $\{\emph{T}_{1}, ... , \emph{T}_{Q}\}$ and (2) \emph{E} = $\{(\emph{T}_{i}, \emph{T}_{j}) |   PC(i,j)  \lor IS (i,j)\}$ where $PC(i,j)$ and $IS (i,j)$ denote that there exist parent-child and immediate-sibling relationship between positions \emph{i} and \emph{j} in the input AST tree respectively.

{The undirected graph is chosen for the following two reasons. On the one hand, the chosen graph structure enables to explore dependencies between transforms applied to different AST nodes in a recursive manner both vertically and horizontally, and the hierarchical nature of AST implies dependencies following these two recursions. On the other hand, the \emph{maximal clique} in the graph is triangle and efficient exact inference algorithm is available for this kind of graph. If the maximal clique in the graph contains more than three nodes, approximate inference algorithms have to be used. \autoref{fig:example} (c) shows the undirected graph for the random field \textbf{T} in \autoref{fig:example} (b). } For the input AST tree in \autoref{fig:example}, there exist parent-child relationship for position pairs <1, 2>, <1, 3>, <3, 4>, <3, 7>, <4, 5>, and <4, 6> and immediate-sibling relationship for position pairs <2, 3>, <4, 7>, and <5, 6>. According to definition 4.2, the graph for random field \textbf{T} is thus as shown in \autoref{fig:example} (c).

The defined undirected graph contains three kinds of cliques: Node clique $\emph{C}_{N}$, Edge clique $\emph{C}_{E}$, and Triangle clique $\emph{C}_{T}$ which contain all vertices, connected edges, and connected triangles in the graph respectively.
In the remaining of this paper, unless explicitly specified, we refer to clique of any kind when we mention clique. 

According to the above definition of random fields \textbf{N} and \textbf{T} and the undirected graph for random field \textbf{T}, we define a CRF over the transforms \textbf{t} to nodes given the observable AST nodes \textbf{n} for the code snippet as:
\begin{equation}
\begin{gathered}
\emph{p}(\textbf{t} | \textbf{n}) = \frac{1}{Z(\textbf{n})} \prod_{{c} \in {C}} \psi_\emph{c}   (\textbf{t}_{c}, \textbf{n})
=\frac{1}{Z(\textbf{n})} \text{exp} \bigg\{ \sum_{c\in \emph{C}}\sum_{k=1}^K \lambda_{k}f_{k}(\textbf{t}_{c}, \textbf{n})\bigg\}
\end{gathered}
\end{equation}
\normalsize

Like our established CRF model, the observable input variables and the predicted output variables in general have the same structure for most applications of CRFs. Thus, each feature function $f_{k}(\textbf{t}_{c}, \textbf{n})$ for a certain clique \emph{c} typically depends on the subset $\textbf{n}_{c}$ of \textbf{n} in the same clique. Considering this, our established CRF model for transform prediction is as follows:
\begin{equation}
\begin{gathered}
\emph{p}(\textbf{t} | \textbf{n}) = \frac{1}{Z(\textbf{n})} \text{exp} \bigg\{ \sum_{c\in \emph{C}}\sum_{k=1}^K \lambda_{k}f_{k}(\textbf{t}_{c}, \textbf{n}_{c})\bigg\} 
\end{gathered}
\end{equation}
\noindent
where $\emph{C} = \emph{C}_{N} \cup \emph{C}_{E} \cup \emph{C}_{T}$ and $\emph{Z}(\textbf{n}) = \sum_{\textbf{t} \in {\Omega}_{\textbf{n}}} \bigg\{\sum_{{c} \in {C}}\sum_{k=1}^K \lambda_{k}f_{k}(\textbf{t}_{c}, \textbf{n}_{c})\bigg\}$.
\normalsize

The feature functions are the key components of CRFs and the learned weight set \{$\lambda_{k}$\} is critical for controlling the probability of a certain  assignment \textbf{t} given the observable \textbf{n}.
For instance, to favor the specific assignment \textbf{t}$_{c}^s$ for a certain clique \emph{c}, a feature function $f_{j}(\textbf{t}_{c}^s, \textbf{n}_{c})$ can be defined with a large numerical value. With the weight $\lambda_{j}$ > 0, the assignment with \textbf{t}$_{c}^s$ for the clique \emph{c} will receive a high conditional probability. 
We will discuss in detail about how feature functions can be defined for our established CRF model in the next section. 

While the idea of structured prediction using CRF in principle can be employed to build a general model that is capable of predicting various kinds of code transforms (\emph{e.g.}, feature improvement, refactoring, and bug-fixing) all at the same time, we in this paper focus on establishing a model for predicting a specific kind of code transform. In other words, the random field \textbf{T} will be only over a certain kind of code transform when we build the model. Building the general model will involve designing more complex feature functions that can effectively separate different kinds of code transforms. 

\subsection{Approach for Structured Prediction of AST-level Code Transform Using CRFs} 
\noindent
After establishment of the CRF model for code transform prediction, we give how our approach can be achieved in this section. 

\noindent
\textbf{Workflow.} Our approach works in two phases. By using the history code transform information $\emph{D}= \{\langle \textbf{t}^{i}, \textbf{n}^{i}\rangle\}_{i=1}^m$ for a set of \emph{m} examples where \textbf{n}$^i$ corresponds to the AST for original code (i.e., before transform) of example \emph{i} and \textbf{t}$^{i}$ corresponds to the transforms applied to nodes of \textbf{n}$^{i}$, we first use an offline training phase to learn a probabilistic model that captures the conditional probability \emph{p}(\textbf{t} | \textbf{n}). Once the model is learned, we then use it to structurally predict the most likely transforms needed for the AST nodes of a new, unseen code snippet. 

We next give a step-by-step process for achieving the approach. We first give the framework for defining code transforms on AST nodes, then illustrate how to define the feature functions used in CRFs, and finally describe the training and prediction process. 

\subsubsection{Framework for Defining AST-level Code Transform}                            
Fine-grained differences between two versions of a code snippet can be described using an edit script which is made up of 
the following 4 basic tree edit operations:

\begin{itemize}
\vspace{0.2mm}
 \item \textbf{UPD}(\emph{x}, \emph{val}): Update the value of a node \emph{x} with the value \emph{val}.

\vspace{0.2mm}
 \item \textbf{ADD}(\emph{x}, \emph{y}, \emph{i}): Add a new node \emph{x}. If the parent \emph{y} is specified, \emph{x} is inserted as the $i^{th}$ child of \emph{y}, otherwise \emph{x} is added as the new root node. 

\vspace{0.2mm}
 \item \textbf{DEL}(\emph{x}): Remove the leaf node \emph{x} from the tree. 

\vspace{0.2mm}
 \item \textbf{MOV}(\emph{x}, \emph{y}, \emph{i}): Move the subtree having node \emph{x} as root to make it the $i^{th}$ child of a parent node \emph{y}.
\end{itemize}

In other words, given two ASTs $ast_{b}$, $ast_{a}$ before and  after code changes, an edit script $\Delta = \{e_{i}|e_{i} \in \{\text{UPD}, \text{ADD}, \text{DEL}, \text{MOV}\} \}$ is a sequence of basic tree edit operations which can convert $ast_{b}$ (AST before change) to $ast_{a}$ (AST after change). We use $|\Delta|$ to denote the number of basic tree edit operations involved for an edit script $\Delta$. 

While code changes typically contain repetitive and predictable patterns \cite{MengPLDI, Prophet}, predicting the edit script itself is typically hard for two reasons. First, a non-trivial code change typically involves multiple tree edit operations and such edit operations can involve complex interactions. Second, predicting a single tree edit operation can involve predicting the edit type, AST nodes involved, and edit position. 
Our insight is that these low-level tree edit operations embody more high-level code transform patterns, and the patterns can be extracted and attached to AST nodes. By lifting the low-level tree edit operations to high-level code transform patterns on AST nodes, powerful structured prediction can be effectively used to establish the relation between code features and AST-level code transform patterns. We below show how to achieve this lifting.

There in general are many possible edit scripts that can achieve the same code changes, an ideal edit script is one with shortest length.  

\noindent \textbf{Definition 4.3.} \textbf{(Shortest Edit Script)}. Let
$\Theta = \{\Delta_{i}|ast_{b} \xrightarrow[\text{}]{\Delta_{i}} ast_{a}\}$ be the set of edit scripts that can convert $ast_{b}$ to $ast_{a}$. An edit script $\Delta_{min} \in \Theta$ is the shortest edit script if  $\forall \Delta^{'} \in (\Theta-\{\Delta_{min}\}), |\Delta^{'}| >= |\Delta_{min}|$. 

The shortest edit script can reflect the essence of code changes, and advanced code differencing tools \cite{gumtree, ChangeDistiller} have been shown effective in producing it. For brevity, hereafter the shortest edit script is directly referred to as edit script.

The edit script can contain too fine-grained tree edit operations, in particular some tree edit operations are related to a high-level AST element but are scattered across the edit script. For example, for the code change that inserts a method call "\emph{call(a)}", there are two add operations: $\text{ADD}(n_{call}, y, i)$ and $\text{ADD}(n_{a}, n_{call}, 1)$, and they are obviously related. Such related but scattered tree edit operations can incur difficulty in extracting high-level concise code transforms. To avoid this issue, we introduce the concept of root tree edit operation. Before the formal definition, we first introduce the concept of mapped node set $N_{m}$. AST differencing algorithms proceed by first  establishing mappings between the similar nodes of $ast_{b}$ and $ast_{a}$. The result of this process includes $N_{m}$ which contains the mapped nodes, along with $N_{o}$ and $N_{n}$ that respectively contain nodes only in $ast_{b}$ and $ast_{a}$. 

\vspace{1.0mm}
\noindent \textbf{Definition 4.4.} \textbf{(Root Tree Edit Operation)}. A basic tree edit operation $\text{O}(x, \_, \_) \in \{\text{UPD}, \text{ADD}, \text{DEL}, \text{MOV}\}$ is a root tree edit operation if (1) O is \text{MOV} operation; or (2) the parent node of \emph{x} belongs to the mapped node set $N_{m}$.

Unlike UPD, ADD, and DEL operations that only
affect one atomic node but not its descendant nodes, MOV operation moves the subtree rooted at one node. Thus, MOV operation can already reflect high-level concise code changes and is always viewed as root tree edit operation. 
For the mentioned two ADD operations $\text{ADD}(n_{call}, y, i)$ and $\text{ADD}(n_{a}, n_{call}, 1)$, 
only $\text{ADD}(n_{call}, y, i)$ is a root tree edit operation as the parent node $n_{call}$ for the other ADD operation belongs to $N_{n}$. 

We now give the definition of code transform on AST node.

\vspace{1.0mm}
\noindent \textbf{Definition 4.5.} \textbf{(Code Transform on AST Node)}. Given the edit script $\Delta$ and a set \emph{O} of root tree edit operations for two ASTs $ast_{b}$, $ast_{a}$ before and after code change, the certain code transform \emph{T} on a certain AST node $\emph{N} \in ast_{b}$ is a predicate $\tt{Tran}$(\emph{T}, \emph{N}): 
\[ \tt{Tran}(\emph{T}, \emph{N})\triangleq \tt{TEO}(\emph{O}, \emph{T}, \emph{N}, \emph{N}^{c}) \land \tt{CON}(\emph{O}, \emph{T}, \emph{N}, \emph{N}^{c})\] 
\noindent 
where $\tt{TEO}(\emph{O}, \emph{T}, \emph{N}, \emph{N}^{c})$ is a predicate about the root tree edit operations on node \emph{N} and node set $\emph{N}^{c}$ that contains nodes that are context related with \emph{N}, and $\tt{CON}(\emph{O}, \emph{T}, \emph{N}, \emph{N}^{c})$ is a predicate about constraints on node $\emph{N}$ and node set $\emph{N}^{c}$. The constraint predicate can be anything related with the AST, like node label, node value, and parent-child relation. There exists code transform $\emph{T}$ on AST node \emph{N} when $\tt{Tran}$(\emph{T}, \emph{N}) evaluates to $\tt{true}$.

If AST node $n \in N_{m}$, let $n^{\mapsto map}$ denote the mapped node of $n$ in the other AST besides the AST that $n$ resides. Context related node is defined as follows. 

\vspace{1.0mm}
\noindent \textbf{Definition 4.6.} \textbf{(Context Related AST node)}. For a certain AST node \emph{N} in AST $ast_{b}$ before code change, another AST node \emph{M} is context related with it if
one of the following dependence relations is satisfied:
\begin{enumerate}
\item (\emph{Data Dependence}) \emph{N} uses or defines a variable whose value is defined in \emph{M};
\item (\emph{Relative Dependence}) \emph{M} is a descendent, ancestor, or sibling of \emph{N};
\item (\emph{Mapping Change Dependence}) If $\emph{N} \in N_{m}$, \emph{M} is $N^{\mapsto map}$ in AST $ast_{a}$ after code change or \emph{M} is a data or relative dependence node for $N^{\mapsto map}$ in $ast_{a}$; If $\emph{N} \notin N_{m}$ but its parent $\emph{p(N)} \in N_{m}$, \emph{M} is $p(N)^{\mapsto map}$ in $ast_{a}$ or \emph{M} is a data or relative dependence node for $p(N)^{\mapsto map}$ in $ast_{a}$.
\end{enumerate}

\subsubsection{Feature Functions for CRFs}
Feature functions are key to controlling the likelihood predictions in CRFs. Similar to the feature functions that have been proven useful in other application areas of CRFs \cite{CRFapplication1, CRFapplication2}, we can consider two types of feature functions for AST-level code transform prediction problem.  

\vspace{1.0mm}
\noindent
\textbf{Observation-based Feature Functions.}
The first type of feature function is called observation-based feature function. For a certain clique \emph{c}, observation-based feature functions typically have the form:
\begin{equation}
\begin{gathered}
f_{k}(\textbf{t}_{c}, \textbf{n}_{c}) = \textbf{1}_{\textbf{t}_{c}=\textbf{t}_{c}'} q_{k} (\textbf{n}_{c}) 
 \nonumber
\end{gathered}
\end{equation}
\normalsize

The notation $\textbf{1}_{\textbf{t}_{c}=\textbf{t}_{c}'}$ is an indicator function of $\textbf{t}_{c}$ which takes the value 1 when $\textbf{t}_{c}=\textbf{t}_{c}'$ and 0 otherwise, $\emph{q} (\textbf{n}_{c})$ is a function on the input $\textbf{n}_{c}$ which we call \emph{observation function}. 
In other words, the feature function is nonzero only for a single output configuration $\textbf{t}_{c}'$. But as long as the constraint is met, then the feature value depends only on the input observation $\textbf{n}_{c}$. Put it in another way, 
we can think of observation-based features as depending only on the input $\textbf{n}_{c}$, but we have a separate set of weights (after learning) for each output configuration. In this case, for a certain clique \emph{c}, we can establish the relation between $\textbf{t}_{c}$ and $\textbf{n}_{c}$ by analyzing the characteristics of input nodes $\textbf{n}_{c}$.

For example, for a node clique ${c}^{,} \in \emph{C}_{N}$ which involves a node $\emph{n}^,$ whose label is method call, we can establish a function $q(\emph{n}^,)$ which analyzes whether the called method has overloaded method(s) and associates the function with different possible transforms that can be applied on this node. Suppose we are focusing on transforms to repair bugs, since overloaded methods are frequently mixed by developers, hopefully the feature function $\textbf{1}_{\textbf{t}_{{c}^{,}}= {\tt Meth-RW-Meth}}q(\emph{n}^,)$ will have a relatively large weight after learning from large data. Same as mentioned in Section 2, here ``{\tt Meth-RW-Meth}'' denotes a repair transform that replaces a method invocation by another one, including overloaded methods. For repair transform, we have designed a set of code element features that are likely to be correlated with repair
transforms on code elements. After learning, the learned weights for the corresponding feature functions will reflect the strength of the correlation. \autoref{tab:type}, \autoref{tab:usage}, and \autoref{tab:syntax} respectively show type related, usage related, and syntax related code element features that we have established. 

\vspace{1.0mm}
\noindent
\textbf{Indicator-based Feature Functions.}
The other type of feature function is called indicator-based feature function, which can be viewed as a pre-processing step before the launch of the learning phase and typically have the following form for a certain clique \emph{c}:
\begin{equation}
\begin{gathered}
f_{j}(\textbf{t}_{c}, \textbf{n}_{c}) = \textbf{1}_{\textbf{t}_{c}=\textbf{t}_{c'}^E \land \textbf{n}_{c}=\textbf{n}_{c'}^E} 
 \nonumber
\end{gathered}
\end{equation}
\normalsize

Here $\textbf{n}_{c'}^E$ and $\textbf{t}_{c'}^E$ denote the input and output for a clique $\emph{c}'$ observed in any of the training data $\emph{D}= \{\langle \textbf{t}^{i}, \textbf{n}^{i}\rangle\}_{i=1}^m$. The idea behind indicator-based feature functions is for each kind of clique $\emph{c}^{T}$ (either node, edge or triangle clique), we get from training data all possible input-output tuples $\langle \textbf{n}_{{c}^{T}}, \textbf{t}_{{c}^{T}}\rangle$ for it and transform each input-output tuple into a feature function. By learning from a large representative dataset, a weight then can be associated with each input-output tuple which in turn can be used to do output predictions for new unseen inputs. 

In summary, indicator-based feature functions are automatically generated from the training set using a domain-
independent procedure. On the contrary, observation-based feature functions can supplement indicator-based feature functions by encoding our domain knowledge about the prediction problem. As previous applications \cite{CRFapplication1, CRFapplication2} of CRFs have shown, these two types of feature functions altogether can deliver good prediction performance. 

\subsubsection{Learning and Prediction} 
Given the training data $\emph{D}= \{\langle \textbf{t}^{i}, \textbf{n}^{i}\rangle\}_{i=1}^m$ of \emph{m} samples, we assume the samples are drawn independently from the underlying joint distribution \emph{p}(\textbf{t}, \textbf{n}) and are identically distributed, i.e., the training data are IID. For our task, we just need to perform discriminative training rather than the more difficult generative training, i.e., we just need to estimate \emph{p}(\textbf{t}|\textbf{n}) directly. The training goal is to automatically compute the optimal weights ${\lambda}$ = \{$\lambda_{k}\}_{k=1}^K$ for feature functions in a way that achieves generalization. In other words, for the set of AST nodes \textbf{n} for a new code snippet drawn from the same distribution \emph{P} (but not contained in the training dataset \emph{D}), its needed code transforms \textbf{t} are predicted correctly by the learned model. There exist several approaches that can potentially be used to accomplish the training task, including for instance \emph{maximum likelihood training} \cite{CRF2} and \emph{max-margin training} \cite{max-margin}.

Based on the above defined feature functions and the learned weight for each feature function, for the AST nodes \textbf{n} of a new code snippet, the conditional probability of each possible transform \textbf{t} can be calculated by substituting the defined feature functions and the learned weights into formula (4) in Section 4.1.

\section{Structured Prediction of Repair Transform Using CRFs}

\noindent
In this section, using repair transform as example, we give a full realization of our approach for structured prediction of code transforms. We first give the definitions of repair transform on AST nodes based on the framework given in Section 4.2.1, then describe how feature functions are constructed for the specific repair transform prediction problem, and finally give the full CRF model learning and prediction algorithms which take the specific issues associated with repair transform prediction into consideration.  

\subsection{Repair Transform on AST Nodes}
\noindent
Using 16 repair transforms as an example, we show how code transforms are attached to AST nodes in this section. Repair transforms are transforms used to change the buggy code into correct code, and are at the heart of many program repair techniques \cite{Kim13, Elixir, Gensis}. The repair transforms used by these existing repair techniques are not at the level of AST node and are in general tried in a fixed order during repair. Our approach instead is able to account the specific characteristics of a certain bug and predict the needed repair transforms at the level of AST node. {The 16 defined repair transforms cover the typical repair actions for the five most common repair patterns \cite{bugpattern,madeiral2018towards}, including change of $\tt{IF}$ condition expression, method call with different actual parameter values, method call with different number of parameters or different types of parameters, change of assignment expression, and addition of $\tt{IF}$ precondition check. Note that our approach itself is not bound to the 16 defined repair transforms, and conceptually it works for any repair transform that would be automatically extracted and attached to AST nodes. Our definition targets object-oriented languages, but can be extended to other languages as well. }

We first give some basic definitions, notations, and utility functions. Based on the definition of AST in Section 3, 
\emph{n}(\emph{lab}, \emph{val}) is used to refer to an AST node with label \emph{lab} and value \emph{val}, $\delta (n)$ and \emph{p(n)} are used to refer to the children nodes and parent node of node \emph{n} respectively, and \emph{l(n)} and \emph{v(n)} are used to refer to the label and value of node \emph{n} respectively. In addition, for a code snippet, $ast^{r}$ is used to refer to the root node of its AST,
and \emph{subast(n)} is used to refer to the sub AST rooted at node \emph{n}. Here, the leaf nodes of the sub AST are always leaf nodes of the original AST. Thus, the sub AST  referred by \emph{subast(n)} is unique for a given code snippet and a certain node \emph{n}. 
Finally, if node \emph{n} is a mapped node produced by the AST differencing algorithm (i.e., $n \in N_{m}$), we use $n^{\mapsto map}$ to denote the mapped node of $n$ in the other AST as in Section 4.

{Other basic definitions, notations, and utility functions are presented in \autoref{fig:notation}. We first give notations for basic tree edit operations (\textbf{AO}) and program elements in mainstream programming languages, including binary operator, logical operator, ternary operator, literal, logical expression, and code block (\textbf{CB}). In particular, the logical expression here denotes an expression made up by a set of atomic boolean expressions, and it cannot be extended with other atomic boolean expressions. We next give some utility mapping functions. For a code snippet, the function \emph{root} achieves the mapping to its root AST node. When the label ${l(n)}$ of an AST node \emph{n} is $\tt{VariableAccess}$ or $\tt{MethodCall}$,
the function \emph{def} maps node \emph{n} to the root node of definition code for ${v(n)}$. 
When the label ${l(n)}$ of an AST node \emph{n} is $\tt{Conditional-If}$, $\tt{LogicalOperator}$ or $\tt{TernaryOperator}$, the function \emph{LogExp} maps node \emph{n} to the root node of the related logical expression. We then give some definitions concerning basic tree edit operations on AST nodes, and the definition is in the from of predicate. \emph{TE}(\emph{e}, \emph{n}) is a predicate about whether there exists tree edit operation \emph{e} on node \emph{n}. Based on predicate \emph{TE}(\emph{e}, \emph{n}), the predicates \emph{NoE}(\emph{n}), \emph{ET}(\emph{n}, \emph{e}), and \emph{NoDE}(\emph{n}) are respectively about whether there exists any tree edit operation on node \emph{n}, whether all nodes of the sub AST rooted at node \emph{n} are subject to tree edit operation \emph{e}, and whether there exists any tree edit operation on related definition code for node \emph{n}. Here, \emph{TE}, \emph{NoE}, \emph{ET}, and \emph{NoDE} stand for the abbreviations for "tree edit", "no edit", "edit tree", and "no definition edit" respectively. We finally give some mapping functions and definitions related with code block. When the label ${l(n)}$ of an AST node \emph{n} is $\tt{Conditional-If}$ or $\tt{Try-Catch}$, the function \emph{block} maps node \emph{n} to the related code block $\textbf{CB}^{l}$ with label \emph{l}. The functions \emph{move} and \emph{min} map a code block to its subset which contains only moved statements and its enclosing statement with the smallest line index respectively. To facilitate the description, \emph{comb}(\emph{n}, \emph{l}) is defined as the combination of functions \emph{min}, \emph{move}, and \emph{block}. }

\begin{figure*}
\centering
\small
\textbf{AO} = \{\text{ADD}, \text{DEL}, \text{MOV}, \text{UPD}\} \quad \quad \quad \, 
\textbf{BinaryOperator} = \{||, \&\&, |, \^ , \&, ==, !=, <, >, <=, >=, <<, >>, +, -, *, /,\% \}

\vspace{0.5mm}
\textbf{LogicalOperator} = \{||, \&\&\} 
\quad \,
\textbf{TernaryOperator} = \{?:\}
\, \textbf{Literal} = \{Number, String, null\} \,  \textbf{CB} = $\{s_1, ..., s_n\}$ 

\vspace{0.5mm}
\textbf{LogicalExpression} = \text{BoolExp}||(\&\&)... ||(\&\&)\text{BoolExp} \, \, \, 
  $ root: \text{code} \to ast^{r} $ \, \, \, 
  $ def: n \to root (v(n)^{\mapsto DEF} )$
  \quad
  
\vspace{0.5mm}
 \emph{LogExp}: $\emph{n} \to root(\emph{n}^{\mapsto \textbf{LogicalExpression}})$ \,\,
 $\emph{TE}(\emph{e},\emph{n}) \triangleq \exists \: e \in \textbf{AO}, \: e(n, \_, \_)$ \,\,
$\emph{NoE}(\emph{n}) \triangleq \forall e \in \textbf{AO}: \neg TE(ao, \emph{$n$}) $   

\vspace{0.5mm}
\footnotesize{
$\emph{ET}(\emph{n}, e) \triangleq 
\forall n' \in subast(n): TE(e,\emph{$n'$})  \quad \quad \quad \,\,\,\, \emph{NoDE}(\emph{n}) \triangleq \forall e \in \textbf{AO} \forall \emph{$n'$} \in \delta (def(n)): \neg TE(e,def(n)) \land  \neg TE(e,\emph{$n'$}) $
}

\vspace{0.5mm}
\emph{block}:$\emph{n} \times \emph{l} \to \textbf{CB}^{l}$ \,
$move:\textbf{CB} \to \{ s \in \textbf{CB} |  \emph{TE}(\text{MOV},root(s))\}$ \,
$min:\textbf{CB} \to \emph{s} ^{min} $ 
$\emph{comb}(\emph{n}, \emph{l})\triangleq \emph{min} \circ \emph{move} \circ \emph{block}(\emph{n}, \emph{l})$ 

\vspace{0.6\baselineskip}   
\caption{Definitions, notations, and utility functions used for defining repair transforms on AST nodes.}
\label{fig:notation}
\end{figure*} 

\begin{table*}
\centering
   \caption{Definitions of the repair transforms that target the inner AST nodes of a statement.}
   \label{tab:definitions-transform-inner} 
   \vspace{-0.6\baselineskip}
\footnotesize
  \begin{tabular}{| p{4.3cm} | p{6.5cm} | p{6.3cm}| }
  \hline
 Repair Transform $\tt{Tran}$ & Predicate About Root Tree Edit Operations $\tt{TEO}$ & Predicate About Constraints $\tt{CON}$ \\
\hline
$\tt{Tran}$ (\tt{Wrap-Meth}, \emph{$n_1^O$}) & $\text{MOV}(\emph{$n_1^O$}, \emph{$n_2^N$}, \emph{$i$}) \land 
\text{ADD}(\emph{$n_2^N$}, \emph{$n_3^O$}, \emph{$j$})\land
 \emph{$NoDE$}(\emph{$n_2^N$}) $
&  $ \emph{l}(\emph{$n_2^N$})=\emph{MethodCall} \land \emph{$n_3^O$} = p(\emph{$n_1^O$})$ \\
\hline
$\tt{Tran}$ (\tt{Unwrap-Meth}, \emph{$n_1^O$}) & $\text{MOV}(\emph{$n_1^O$}, \emph{$n_2^O$}, \emph{i}) \land 
\text{DEL}(\emph{$n_3^O$}) \land
 \emph{NoDE}(\emph{$n_3^O$}) $
&  $ \emph{l}(\emph{$n_3^O$})=\emph{MethodCall} \land \emph{$n_3^O$} = p(\emph{$n_1^O$}) \land  \emph{$n_2^O$} = p(\emph{$n_3^O$})$ \\
\hline
$\tt{Tran}$ (\tt{Var-RW-Var},\emph{$n_1^O$}) &  $ \text{UPD}(\emph{$n_1^O$}, val) \land
 \emph{$NoDE$}(\emph{$n_1^O$}) \land \emph{$NoDE$}\big(\emph{$(n_1^O)^{\mapsto map}$}\big) $ 
& $\emph{l}(\emph{$n_1^O$})=\emph{VariableAccess}$ \\
\hline
$\tt{Tran}$ (\tt{Var-RW-Meth},\emph{$n_1^O$}) &  $
\text{DEL}(\emph{$n_1^O$}) \land 
\text{ADD}(\emph{$n_2^N$}, \emph{$n_3^O$}, \emph{j}) \land
\emph{NoDE}(\emph{$n_1^O$}) \land \emph{NoDE}(\emph{$n_2^N$}) $ 
& $\emph{l}(\emph{$n_1^O$})=\emph{VariableAccess} \land \emph{l}(\emph{$n_2^N$})=\emph{MethodCall} \land \emph{$n_3^O$} = p(\emph{$n_1^O$}) $ \\
\hline
$\tt{Tran}^1$(\tt{Meth-RW-Meth},\emph{$n_1^O$}) &  $ \text{UPD}(\emph{$n_1^O$}, val) \land
 \emph{$NoDE$}(\emph{$n_1^O$}) \land \emph{$NoDE$}\big(\emph{$(n_1^O)^{\mapsto map}$}\big)  $ 
& $\emph{l}(\emph{$n_1^O$})=\emph{MethodCall}$  \\
\hline
$\tt{Tran}^2$(\tt{Meth-RW-Meth},\emph{$n_2^O$}) &  $ \text{DEL}(\emph{$n_1^O$}) \land 
\emph{NoDE}(\emph{$n_2^O$})$ & 
$ \emph{$n_2^O$} = p(\emph{$n_1^O$}) \land \emph{l}(\emph{$n_2^O$})=\emph{MethodCall}$  \\
\hline
$\tt{Tran}$ (\tt{Meth-RW-Var},\emph{$n_1^O$}) &  $
\text{DEL}(\emph{$n_1^O$}) \land 
\text{ADD}(\emph{$n_2^N$}, \emph{$n_3^O$}, \emph{j}) \land
\emph{NoDE}(\emph{$n_1^O$}) \land \emph{NoDE}(\emph{$n_2^N$}) $ 
& $\emph{l}(\emph{$n_1^O$})=\emph{MethodCall} \land \emph{l}(\emph{$n_2^N$})=\emph{VariableAccess} \land \emph{$n_3^O$} = p(\emph{$n_1^O$}) $ \\
\hline
$\tt{Tran}$(\tt{BinOper-Rep},\emph{$n_1^O$}) &  $  
\text{UPD}(\emph{$n_1^O$}, val) \land \forall \emph{$n'$} \in \delta({n_{1}^O}): \emph{NoE}(\emph{$n'$}) $ 
& $\emph{l}(\emph{$n_1^O$})=\emph{BinaryOperator} $ \\
\hline
$\tt{Tran}$ (\tt{Constant-Rep}, \emph{$n_1^O$}) &  $  
\text{UPD}(\emph{$n_1^O$}, val) \land \emph{NoE}(p(\emph{$n_1^O$})) $ 
& $\emph{l}(\emph{$n_1^O$})=\emph{Literal} $ \\
\hline
$\tt{Tran}$ (\tt{LogExp-Exp}, \emph{$LogExp$($n_2^O$})) &  $  
\text{ADD}(\emph{$n_1^N$}, \emph{$n_2^O$}, \emph{i}) \land \exists \emph{$n'$} \in \delta({n_{1}^N}): \emph{NoE}(\emph{$n'$}) \land 
\exists \emph{$n''$} \in \delta({n_1^N}):  \emph{ET}(\emph{$n''$}, \text{ADD}) $ & $\emph{l}(\emph{$n_1^N$})=\emph{LogicalOperator}
$ \\
\hline
$\tt{Tran}$ (\tt{LogExp-Red}, \emph{$LogExp$($n_1^O$})) &  $  
\text{DEL}(\emph{$n_1^O$}) \land \exists \emph{$n'$} \in \delta({n_{1}^O}): \emph{NoE}(\emph{$n'$}) \land 
\exists \emph{$n''$} \in \delta({n_1^O}):  \emph{ET}(\emph{$n''$}, \text{DEL}) $ & $\emph{l}(\emph{$n_1^O$})=\emph{LogicalOperator}
$ \\
\hline
\end{tabular}
\end{table*}

\begin{table*}
\centering
  \footnotesize
  \caption{Definitions of the repair transforms that target majorly the virtual root node of a statement. }
   \label{tab:definitions-transform-virtual} 
   \vspace{-0.6\baselineskip}
\footnotesize
  \begin{tabular}{| p{4.3cm} | p{6.5cm} | p{6.3cm}| }
  \hline
 Repair Transform $\tt{Tran}$ & Predicate About Root Tree Edit Operations $\tt{TEO}$ & Predicate About Constraints $\tt{CON}$ \\
\hline
$\tt{Tran}$ (\tt{Wrap-IF-N}, $\emph{$comb$($n_1^N$,THEN})^{\mapsto vr \mapsto map}$) &  $ 
\text{ADD}(\emph{$n_1^N$}, \emph{$n_2^O$}, \emph{i}) \land  \emph{move}\big(\emph{block}(\emph{$n_1^N$}, THEN)\big) != \emptyset $
 & $\emph{l}(\emph{$n_1^N$})=\emph{Cond If} \land \emph{block}(\emph{$n_1^N$}, ELSE) = \emptyset \land n(Literal, NULL) \in subast\Big(root\big(LogExp(\emph{$n_1^N$})\big)\Big) $\\
\hline
$\tt{Tran}^1$ (\tt{Wrap-IFELSE-N}, $\emph{$comb$($n_1^N$,THEN})
^{\mapsto vr \mapsto map}$)
&  $ 
\text{ADD}(\emph{$n_1^N$}, \emph{$n_2^O$}, \emph{i}) \land  \emph{move}\big(\emph{block}(\emph{$n_1^N$}, THEN)\big) != \emptyset \land  \emph{move}\big(\emph{block}(\emph{$n_1^N$}, ELSE)\big) = \emptyset $
 & $\emph{l}(\emph{$n_1^N$})=\emph{Cond If} \land \emph{block}(\emph{$n_1^N$}, ELSE) != \emptyset \land n(Literal, NULL) \in subast\Big(root\big(LogExp(\emph{$n_1^N$})\big)\Big) $\\
\hline
$\tt{Tran}^2$ (\tt{Wrap-IFELSE-N}, $\emph{$comb$($n_1^N$,ELSE})
^{\mapsto vr \mapsto map}$) &  $ 
\text{ADD}(\emph{$n_1^N$}, \emph{$n_2^O$}, \emph{i}) \land  \emph{move}\big(\emph{block}(\emph{$n_1^N$}, THEN)\big) = \emptyset \land  \emph{move}\big(\emph{block}(\emph{$n_1^N$}, ELSE)\big) != \emptyset $
 & $\emph{l}(\emph{$n_1^N$})=\emph{Cond If} \land \emph{block}(\emph{$n_1^N$}, THEN) != \emptyset \land n(Literal, NULL) \in subast\Big(root\big(LogExp(\emph{$n_1^N$})\big)\Big) $\\
\hline
$\tt{Tran}^3$ (\tt{Wrap-IFELSE-N}, \emph{$n_1^O$}) &  $\text{MOV}(\emph{$n_1^O$}, \emph{$n_2^N$}, \emph{i}) \land 
\text{ADD}(\emph{$n_2^N$}, \emph{$n_3^O$}, \emph{j}) $
 & $ \emph{l}(\emph{$n_2^N$})=\emph{TernaryOperator} \land \emph{$n_3^O$} = p(\emph{$n_1^O$}) \land n(Literal, NULL) \in subast\Big(root\big(LogExp(\emph{$n_2^N$})\big)\Big) $ \\
\hline
$\tt{Tran}^1$ (\tt{Unwrap-IF},\emph{$n_1^O$}) &  $ \text{DEL}(\emph{$n_1^O$}) \land 
\emph{move}\big(\emph{block}(\emph{$n_1^O$}, THEN)\big) \cup \emph{move}\big(\emph{block}(\emph{$n_1^O$}, ELSE)\big) != \emptyset $
 & $ \emph{l}(\emph{$n_1^O$})=\emph{Cond If} $ \\
\hline
$\tt{Tran}^2$(\tt{Unwrap-IF}, \emph{$n_1^O$}) &  $ \text{DEL}(\emph{$n_1^O$}) \land \text{MOV}(\emph{$n_2^O$}, \emph{$n_3^O$}, \emph{i}) $
 & $ \emph{l}(\emph{$n_1^O$})=\emph{TernaryOperator} \land \emph{$n_1^O$} = p(\emph{$n_2^O$}) \land \emph{$n_3^O$} = p(\emph{$n_1^O$})  $ \\
\hline
$\tt{Tran}$ (\tt{Wrap-TRY}, $\emph{$comb$($n_1^N$,TRY})
^{\mapsto vr \mapsto map}$)
&  $ \text{ADD}(\emph{$n_1^N$}, \emph{$n_2^O$}, \emph{i}) \land  \emph{move}\big(\emph{block}(\emph{$n_1^N$}, TRY)\big) != \emptyset $
 & $\emph{l}(\emph{$n_1^N$})=\emph{Try} 
$ \\
\hline
\end{tabular}
\end{table*}

\begin{figure*}
\centering
\makebox[0pt]{
 \includegraphics[scale=0.208]{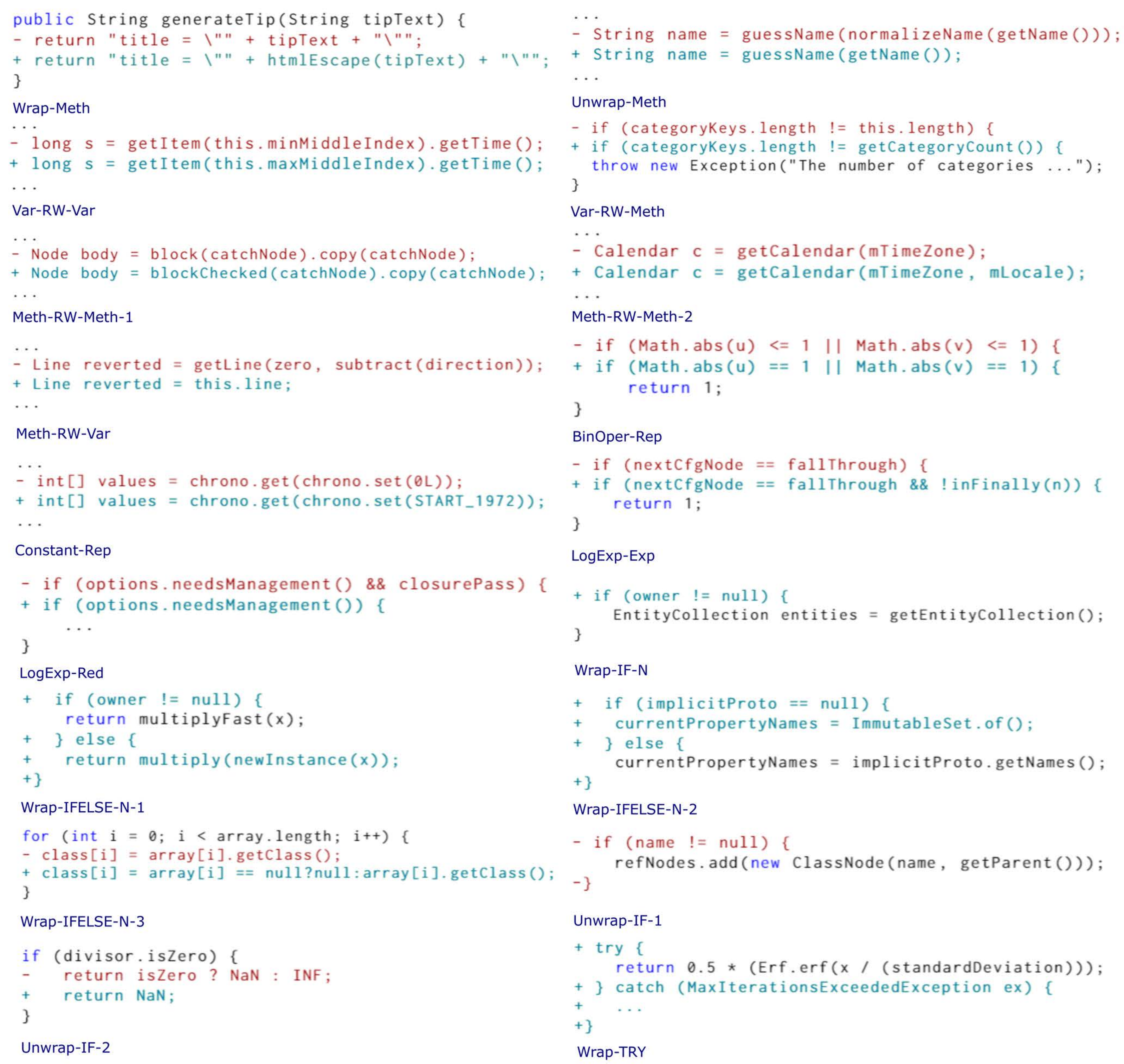}}
 \caption{
  Examples to illustrate the definitions of repair transforms.
}
 \label{fig:transformexample}
\end{figure*}
\noindent

{\autoref{tab:definitions-transform-inner} and \autoref{tab:definitions-transform-virtual} give the definitions for 16 repair transforms in detail, and \autoref{fig:transformexample} uses examples to illustrate the repair transforms corresponding to the definitions in the two tables. 
For the definitions, the superscripts `O' and `N' are used to distinguish nodes in ASTs before and after code change respectively, the symbols `Wrap', `Unwrap', and `RW' in transform name are used to denote that an existing program element is wrapped with other newly added constructs (e.g., method call, conditional check), an existing program element is unwrapped from other existing constructs, and an existing program element is replaced with other program elements respectively. 
}

\autoref{tab:definitions-transform-inner} gives the definitions for some repair transforms that target the inner nodes of a statement. $\tt{Wrap-Meth}$ moves an existing expression into an added method call and $\tt{Unwrap-Meth}$ instead moves an existing expression out of a removed method call. {To avoid the case that the move operations arise because of changes in the signature of the involved method call, we add a constraint that there are no root tree edit operations on the root node of the method definition. Note that when the definition explicitly involves a certain variable access or method call, we in general have constraints about the tree edit operations on the definitions of the accessed variable or method. }$\tt{Var-RW-Var}$, $\tt{Var-RW-Meth}$, $\tt{Meth-RW-Meth}$, and $\tt{Meth-RW-Var}$ achieve the replacement of variable access and method call. In particular, there are two sub-cases for $\tt{Meth-RW-Meth}$: the names of the replaced and original method calls are (1) different or (2) the same (i.e., method overload). $\tt{BinOper-Rep}$ and $\tt{Constant-Rep}$ represent the replacement of binary operator and constant literal respectively. $\tt{LogExp-Exp}$ and $\tt{LogExp-Red}$ expand and reduce an existing logical expression respectively. { As logical expression can typically be expanded in different ways when it contains several atomic boolean expressions, we thus associate both these two repair transforms to the root node of the logical expression.}

Other transforms target majorly the whole statement and are given in \autoref{tab:definitions-transform-virtual}. $\tt{{Wrap-IF-N}}$ adds an `If' conditional check (without `Else' block) for an existing statement that is subject to MOV operation. {Note that some other tree edit operations on the `Then' block can be accompanied by the add of the `If' conditional check. In such cases, we view the first statement in the `Then' block (whose actual root node is subject to MOV operation) as the `old' statement and the target of the conditional check.} $\tt{Wrap-IFELSE-N}$ adds an `If' conditional check with the `Else' block. For this transform, we have 3 cases: the `old' (moved) statement is wrapped by (1) `Then' block, or (2) `Else' block, or (3) the added check is in the form of ternary expression. For both $\tt{{Wrap-IF-N}}$ and $\tt{Wrap-IFELSE-N}$, 
the logical condition of the added `If' check has the node $n(Literal, NULL)$ (i.e., the added check is null check). Similarly, we consider additional transforms $\tt{{Wrap-IF-O}}$ and $\tt{{Wrap-IFLSE-O}}$ (not shown in \autoref{tab:definitions-transform-virtual} for space reason) for which the added `If' check is not related to null check. $\tt{Unwrap-IF}$ removes the conditional check, and the check can be in the form of `If' expression or ternary expression. Finally, $\tt{Wrap-TRY}$  warps an existing statement with "\emph{Try-Catch}" exception handle. 

For "\emph{Wrap-If}" and "\emph{Wrap-Try}" related transforms that target the whole statement \emph{s}, we introduce the "virtual root" node (denoted as $s^{\mapsto vr}$) and attach transforms to it rather than the actual root node $root(s)$. 
The aim is to separate other transforms that can possibly be attached to $root(s)$. For instance, for a method invocation statement whose actual root node is the called method, there can exist both $\tt{{Wrap-IF-N}}$ transform and $\tt{Meth-RW-Meth}$ transform that replaces the called method, and  $\tt{Meth-RW-Meth}$ transform is attached to the actual root node. The "virtual root" node $s^{\mapsto vr}$ is inserted between the actual root node $root(s)$ and its parent node. Doing this also facilities the CRF model construction as it typically has a single-label per input assumption, i.e., single repair transform per AST node for our problem. For "virtual root" node $s^{\mapsto vr}$, we view the label and value of it as `virtual' and `null' respectively. 
\vspace{1mm}

\noindent 
\textbf{Example} For the code diff in \autoref{fig:overview}, the edit script consists of two root tree edit operations 
\textbf{UPD}(\circled{\scriptsize 11}, \emph{z}) and \textbf{UPD}(\circled{\scriptsize 13}, \emph{z}) where \circled{\scriptsize 11} and \circled{\scriptsize 13} denote the 11th and 13th AST node respectively (as shown in \autoref{fig:overview}). Since the definitions for the old method \emph{y} and the new method \emph{z} have not changed, the predicates \emph{$NoDE$}(\emph{$\circled{\scriptsize 11}^O$}) and \emph{$NoDE$}\big(\emph{$(\circled{\scriptsize 11}^O)^{\mapsto map}$}\big) evaluate to true (the part of predicate $\tt{TEO}$ about root tree edit operations). Meanwhile, the predicate "$\emph{l}(\emph{$\circled{\scriptsize 11}^O$})=\emph{MethodCall}$" also evaluates to true (the part of predicate $\tt{CON}$ about constraints). Consequently, the predicate $\tt{Tran}^1$($\tt{Meth-RW-Meth}$, \emph{$\circled{\scriptsize 11}^O$}) evaluates to true according to the definition given in \autoref{tab:definitions-transform-inner}, implying that the repair transform "{\tt Meth-RW-Meth}" is attached to node \circled{\scriptsize 11}. Similarly, we can establish that 
the repair transform "{\tt Meth-RW-Meth}" is also attached to 
node \circled{\scriptsize 13}.

\subsection{Feature Functions for Repair Transform Prediction}
\noindent
We in this section describe how observation-based and indicator-based feature functions can be  constructed for the repair transform prediction problem.

\subsubsection{Observation-based Feature Functions}
For a certain clique \emph{c}, observation-based feature functions establish the relation between $\textbf{t}_{c}$ and $\textbf{n}_{c}$ by analyzing the characteristics of input nodes $\textbf{n}_{c}$. With regard to repair transform, this basically means designing a set of node characteristics that are likely to be correlated with repair transforms on them. After learning, the learned weights for the corresponding feature functions then reflect the strength of the correlation. 

We design observation-based feature functions related with different kinds of program elements reflected in AST nodes, including variable access, method call, logical expression, binary operator, and the whole statement. We first present the characteristics that we analyze and then present the observation-based feature functions on top of them. 

\noindent \textbf{Node Characteristics.}
Depending on the label of the AST node, we accordingly analyze different kinds of characteristics associated with it and the characteristics we statically analyze can be classified into 3 kinds based on their nature: type related, usage related, and syntax related. 

\begin{table*}
\centering
 \caption{Type related node characteristics.}
\begin{adjustwidth}{-1.5in}{-1.5in}
  \small
   \vspace{-0.6\baselineskip}
  \label{tab:type}
  \centering
  \begin{tabular}{|p{1.6cm}|p{16.0cm}|}
  \hline
  Node Kind & Characteristic Description\\
  \hline
    \hline
    \multirowcell{7}{Variable \\ Access \\ \scriptsize(\emph{var})}  & ($\tt{V1}$): The type of \emph{var} is primitive. \\ 
    \cline{2-2}
    & ($\tt{V2}$): The type of \emph{var} is objective. \\
    \cline{2-2}
    & ($\tt{V3}$): \emph{var} is an instance of the class that it resides.\\
    \cline{2-2}
     & ($\tt{V4}$): There exist variables in scope that are type compatible with \emph{var}.\\
     \cline{2-2}
     & ($\tt{V5}$): There exist method definitions or calls for which at least one of their parameters is type compatible with \emph{var}.\\
     \cline{2-2}
     & ($\tt{V6}$): There exist method definitions or calls whose return types are compatible with \emph{var}.\\
    \hline
    \multirowcell{5}{Method \\ Invocation \\ \scriptsize(\emph{m})}  & ($\tt{M1}$): The return type of \emph{m} is primitive. \\ 
    \cline{2-2}
    & ($\tt{M2}$): The return type of \emph{m} is objective. \\
    \cline{2-2}
    & ($\tt{M3}$): The types of some parameters of \emph{m} are compatible with the return type of \emph{m}.\\
    \cline{2-2}
     & ($\tt{M4}$): There exist variables in scope that are type compatible with the return type of \emph{m}. \\
     \cline{2-2}
     & ($\tt{M5}$): There exist method definitions or calls whose return types are compatible with that of \emph{m}.\\
    \hline
\end{tabular}
   \vspace{-0.8\baselineskip}
       \end{adjustwidth}
\end{table*}

\begin{table*}
 \caption{Usage related node characteristics.}
\begin{adjustwidth}{-1.5in}{-1.5in}
  \small
   \vspace{-0.6\baselineskip}
  \label{tab:usage}
  \centering
  \begin{tabular}{|p{1.6cm}|p{16.0cm}|}
  \hline
  Node Kind & Characteristic Description\\
  \hline
    \hline
    \multirowcell{13}{Variable \\ Access \\ \scriptsize(\emph{var})}  & ($\tt{V7}$): 
    If \emph{var} is a local variable, it has not been referenced before the statement that \emph{var} resides.
    \\ \cline{2-2}
    & ($\tt{V8}$): If \emph{var} is a local variable, it has not been assigned before the statement that \emph{var} resides.
    \\ \cline{2-2}
    & ($\tt{V9}$): 
    If \emph{var} is a field, it has not been referenced in other statements of the class besides the statement that \emph{var} resides.
    \\ \cline{2-2}
     & ($\tt{V10}$): If \emph{var} is a field, it has not been assigned in other statements of the class besides the statement that \emph{var} resides.
     \\  \cline{2-2}
     & ($\tt{V11}$): There exist other statements in the class that use some same type variables with \emph{var}, but have null check guard.
     \\ \cline{2-2}
     & ($\tt{V12}$): There exist other statements in the class that use some same type variables with \emph{var}, but have normal check guard.
     \\ \cline{2-2}
    & ($\tt{V13}$): When \emph{var} is a parameter of a method call $\emph{m}_{1}$, replace \emph{var} with another variable \emph{${var}'$} can get another method call $\emph{m}_{2}$ used in the class.  
     \\ \cline{2-2}
    & ($\tt{V14}$): When \emph{var} is a parameter of a method call $\emph{m}_{1}$, replace \emph{var} with a method call \emph{${m}'$} can get another method call $\emph{m}_{2}$ used in the class. \\
    \hline
    \multirowcell{9}{Method \\ Invocation \\ \scriptsize(\emph{m})}  & ($\tt{M6}$): There exist other statements in the class that use a method call $\emph{m}'$ whose signature is same with \emph{m}, but have null check guard.
    \\   \cline{2-2}
    & ($\tt{M7}$): There exist other statements in the class that use a method call $\emph{m}'$ whose signature is same with  \emph{m}, but have normal check guard.
    \\   \cline{2-2}
    & ($\tt{M8}$): There exist other statements in the class that use a method call $\emph{m}'$ whose signature is same with \emph{m}, but have try-catch exception handle.
    \\  \cline{2-2}
     & ($\tt{M9}$): When \emph{m} is a parameter of a method call $\emph{m}_{1}$, replace \emph{m} with a variable \emph{var} can get another method call $\emph{m}_{2}$ used in the class.
     \\   \cline{2-2}
     & ($\tt{M10}$): When \emph{m} is a parameter of a method call $\emph{m}_{1}$, replace \emph{m} with a method call \emph{${m}'$} can get another method call $\emph{m}_{2}$ used in the class. \\
    \hline
\end{tabular}
   \vspace{-0.8\baselineskip}
       \end{adjustwidth}
\end{table*}

\begin{table*}
 \caption{Syntax related node characteristics.}
    \begin{adjustwidth}{-1.5in}{-1.5in}
  \small
   \vspace{-0.6\baselineskip}
  \label{tab:syntax}
  \centering
  \begin{tabular}{|p{1.6cm}|p{16.0cm}|}
  \hline
  Node Kind & Characteristic Description\\
  \hline
    \hline
    \multirowcell{2}{Variable \\ Access \scriptsize (\emph{var})}  & ($\tt{V15}$): There exist variables in scope that are similar in identifier name with \emph{var}.
    \\ \cline{2-2}
    & ($\tt{V16}$): There exist method definitions or calls that are similar in identifier name with \emph{var}.
    \\  \hline
    \multirowcell{4}{Method \\ Invocation \\ \scriptsize(\emph{m})}  & ($\tt{M11}$): The identifier name of \emph{m} starts with `get'.
    \\ \cline{2-2}
     & ($\tt{M12}$): \emph{m} has overloaded method(s).
     \\  \cline{2-2}
     & ($\tt{M13}$): There exist variables in scope that are similar in identifier name with \emph{m}. 
     \\ \cline{2-2}
     & ($\tt{M14}$): There exist method definitions or calls that are similar in identifier name with \emph{m}. \\
    \hline
     \multirowcell{4}{Binary \\ Operator \\ \scriptsize(\emph{bo})}  & ($\tt{BO1}$): The operator kind of \emph{bo}.
     \\  \cline{2-2}
    & ($\tt{BO2}$): When \emph{bo} is a logical operator, its operands contain the exclamation mark!.
    \\  \cline{2-2}
    & ($\tt{BO3}$): When \emph{bo} is a logical operator, its operands contain the literal `null'.
    \\  \cline{2-2}
     & ($\tt{BO4}$): When \emph{bo} is a logical operator, its operands contain the number `0' or `1'.\\
    \hline
     \multirowcell{3}{Root  Node\\ }  & ($\tt{LE1}$): There exists an atomic boolean expression that contains the exclamation mark !.
     \\  \cline{2-2}
    & ($\tt{LE2}$): There exists an atomic boolean expression that is simply a boolean variable. 
    \\  \cline{2-2}
     & ($\tt{LE3}$): There exist two atomic boolean expressions $\emph{ae}_{1}$ and $\emph{ae}_{2}$ that are respectively null check and normal check. \\
    \hline
     \multirowcell{5}{Virtual \\ Root Node}  & ($\tt{S1}$): The statement kind of \emph{s}.
     \\ \cline{2-2}
    & ($\tt{S2}$): The statement kind of the previous statement in the same block with \emph{s}.
    \\  \cline{2-2}
     & ($\tt{S3}$): The statement kind of the next statement in the same block with \emph{s}.
     \\ \cline{2-2}
      & ($\tt{S4}$): The statement kind of the parent statement of \emph{s}.
      \\ \cline{2-2}
      & ($\tt{S5}$): The associated method of \emph{s} throws exception or the associated class of \emph{s} extends an exception type class.\\
    \hline
\end{tabular}
   \vspace{-0.8\baselineskip}
       \end{adjustwidth}
\end{table*}

{\autoref{tab:type}, \autoref{tab:usage}, and \autoref{tab:syntax} respectively show type related, usage related, and syntax related node characteristics that we have established. For type related characteristics, we majorly explore whether the type is primitive or objective and whether certain type compatible relation holds. We investigate types of different kinds of program elements, including variables in different forms (local variable, method parameter, etc.) and returns of method definitions or method calls. In addition, we also examine whether types of certain program elements (regardless of same kind or different kind) are compatible with each other when the type comparison is applicable. For usage related characteristics, we consider how the variable or method call has been used elsewhere in the program and whether certain substitutions of variable or method call can result in other program elements in the program. We explore whether variable or method call has been used elsewhere with null check guard, normal check guard, and exception handle. When they are used as a method call parameter, we explore whether substitution between different variables (resp. different method calls) or substitution between variable and method call can result in another method call. For variable, we additionally investigate whether it has been referenced or assigned elsewhere and we distinguish local variable from field. For syntax related characteristics, we explore whether certain program elements have some specific syntax attributes. With regard to variable and method call, we majorly investigate whether their  identifier names are special, including similar with that of others or start with the special character `get'. For binary operator and root node for logical expression, we primarily study whether their children are unusual, including whether or not contain the exclamation mark !, the literal `null', the number `0', or the number `1'. In terms of virtual root node for a certain statement \emph{s}, we mainly explore the statement kinds of several statements related with \emph{s}, including the statement \emph{s} itself, previous statement of \emph{s}, next statement of \emph{s}, and parent statement of \emph{s}. }

When the characteristic involves measure of similarity ($\tt{V15}$, $\tt{V16}$, $\tt{M13}$, and $\tt{M14}$), we use Levenshtein distance metric to establish the difference between two string sequences. For the characteristics related with statement kind ($\tt{S1}$-$\tt{S4}$) or binary operator kind ($\tt{BO1}$), we enumerate all possible kinds and establish a sub-characteristic \emph{"The kind is X"} for each possible kind \emph{X}. Typical binary operator kinds in mainstream languages include logical relation, bit operation, equality comparison, shift operation, and mathematical operation as shown in \autoref{fig:notation}. After doing this, each of the characteristics can be viewed as a boolean valued function, i.e, a predicate on the characteristics. We use $\emph{n.c}$ to denote the boolean evaluation result of a certain characteristic \emph{c} on node \emph{n}.

\emph{Characteristic Propagation.} As there exist structural dependencies between different AST nodes, the characteristic of a certain node can possibly imply repair transforms on other nodes. For instance, the characteristic $\tt{V7}$ 
can be related with \emph{"Wrap-If"} related transform(s) on the virtual root node of the statement. To account this, we propagate some characteristics of certain child nodes to their parent nodes. First, we propagate the characteristics $\tt{V7}$, $\tt{V8}$, $\tt{V9}$, $\tt{V10}$, $\tt{V11}$, and $\tt{V12}$ for a 
variable access node $\emph{n}_{1}$ to the virtual root node of the statement and the method access node $\emph{n}_{2}$ that is an ancestor of $\emph{n}_{1}$. Second, we propagate the characteristics $\tt{M6}$, $\tt{M7}$, and $\tt{M8}$ 
for a method access node to the virtual root node of the statement. When propagating a certain characteristics \emph{c} from a node $\emph{n}_{1}$ to another node $\emph{n}_{2}$, the corresponding characteristic $\emph{c}^{,}$ for $\emph{n}_{2}$ is "There exists at least one descendent node that has characteristic \emph{c}", and the predicate value is calculated as follows:  
\begin{equation}
\begin{gathered}
 {n}_{2}.{c}' =  {n}_{c1}.{c} \lor
 {n}_{c2}.{c} \lor ... \lor{n}_{ck}.{c}
 \nonumber
\end{gathered}
\end{equation} 
\normalsize
\noindent where ${n}_{c1}$ to ${n}_{ck}$ represent \emph{k} descendent nodes of $\emph{n}_{2}$ that have characteristic \emph{c}, including $\emph{n}_{1}$. 

\noindent \textbf{Observation-based Feature Functions.}
After analyzing the characteristics, the observation function $\emph{q} (\textbf{n}_{c})$ can then be designed as indicator function of the characteristics.
Let $\emph{C}^L= \{\emph{c}^i\}_{i=1}^n$ denote the set of characteristics we have established for node whose label is \emph{L}. For a node \emph{n} and a characteristic $\emph{c} \in \emph{C}^{L}$, we define the observation function 
as follows:
\begin{equation}
\begin{gathered}
\emph{q} (n, c) = \textbf{1}_{l(n) = L \land \emph{n.c} = true}  
 \nonumber
\end{gathered}
\end{equation}  
\normalsize

Then, we need to correlate the observation function with repair transforms. For different types of nodes, the set of possible repair transforms on them are different. To see what transforms are possible for a certain node, we make use of information from the training data $\emph{D}= \{\langle \textbf{t}^i, \textbf{n}^i\rangle\}_{i=1}^m$ of \emph{m} samples. For repair transform prediction problem, the training data $\emph{D}= \{\langle \textbf{t}^i, \textbf{n}^i\rangle\}_{i=1}^m$ consists of \emph{m} buggy programs where for each program, the repair transforms required to fix the bug are associated to appropriate AST nodes. For an AST node $n \in \textbf{n}$, we use \emph{t(n)} to denote the associated repair transform on it, and we define the viable transform set for a node whose label is \emph{L} as follows:
\[\emph{T} (L) = \{ t | \exists \langle \textbf{t}, \textbf{n} \rangle \in \emph{D}, \: \exists n \in \textbf{n} : l(n) = L, t(n)=t \} 
\nonumber
\]
\normalsize

That is, we deem a repair transform possible for a certain type of node when we have observed the occurrence of this at least once in the training data. 
Let $\gamma$ represent the set of possible labels for a node, we finally define observation-based feature functions on top of the observation functions and the viable transform set as follows:
 \begin{equation}
\begin{gathered}
f(t, n) = \textbf{1}_{L \in \gamma \land l(n)=L \land t \in \emph{T} (L) \land c \in \emph{C}^L} \emph{q} (n, c)
\nonumber
\end{gathered}
\end{equation}
\normalsize

Intuitively speaking, for each possible repair transform \emph{t} on a node whose label is \emph{L}, we associate it with each of the characteristics we have designed for nodes with label \emph{L} to form a feature function. Through training on big data, we can get weights for different transform-characteristic pairs. 

Note that we in this paper define observation-based feature functions on node cliques, and it is possible to define more complex observation-based feature functions on edge cliques and triangle cliques by analyzing the characteristics involving all nodes in edge and triangle cliques.  

\subsubsection{Indicator-based Feature Functions}
\label{sec:indicator}
Given the training data $\emph{D}= \{\langle \textbf{t}^{i}, \textbf{n}^{i}\rangle\}_{i=1}^m$ of \emph{m} sample repair examples, we now formally describe how we establish indicator-based feature functions. Recall that \emph{l(n)}, \emph{v(n)}, and \emph{t(n)} are used to denote node label, node value, and repair transform on node respectively. Here, we also use $n^{\mapsto i}$ to represent the \emph{i}th child node of node \emph{n}. For each tuple $\langle \textbf{t}, \textbf{n} \rangle \in \emph{D}$, we define the observed set of transforms on nodes for different kinds of cliques as follows:

\begin{equation}
\begin{split}
 nodetran (\textbf{t}, \textbf{n}) = \{(T,L) | \exists \emph{n} \in \textbf{n}: l(n)=L, t(n)=T\} \\
& 
\nonumber
\end{split}
\end{equation}
\vspace{-7mm}
\begin{equation}
\begin{split}
 edgetran (\textbf{t}, \textbf{n}) = \{(T_1, L_1,T_2, L_2) | \exists \emph{n} \in \textbf{n}, \emph{i} \in \mathbb N: l(n)=L_1, \\ t(n)=T_1, l(n^{\mapsto i})=L_2, t(n^{\mapsto i})=T_2\} \\
& 
\nonumber
\end{split}
\end{equation}
\vspace{-7mm}
\begin{equation}
\begin{split}
 triangletran (\textbf{t}, \textbf{n}) = \{(T_1, L_1,T_2,L_2,T_3, L_3) | \exists \emph{n} \in \textbf{n}, \emph{i} \in \mathbb N: \\ l(n)=L_1,  t(n)=T_1, l(n^{\mapsto i})=L_2, t(n^{\mapsto i})=T_2, \\ l(n^{\mapsto (i+1)})=L_3, t(n^{\mapsto (i+1)})=T_3\}
\nonumber
\end{split}
\end{equation}

In other words, we enumerate possible repair transforms on nodes for different cliques observed in the specific training example $\langle \textbf{t}, \textbf{n} \rangle$. 
For the entire training data \emph{D}, we can then obtain all observed set of transforms on nodes as follows:
\begin{equation}
\begin{gathered}
all\_nodetran (D) = \cup_{i=1}^m nodetran (\textbf{t}^i, \textbf{n}^i) \\[0pt]
all\_edgetran (D) = \cup_{i=1}^m edgetran (\textbf{t}^i, \textbf{n}^i) \\[0pt]
all\_triangletran (D) = \cup_{i=1}^m triangletran (\textbf{t}^i, \textbf{n}^i) \nonumber
\end{gathered}
\end{equation}
\normalsize

Based on the observed set of transforms on nodes, we finally define indicator-based feature functions for different kinds of cliques as follows: 
\begin{equation}
\begin{gathered}
f({t}_{1}, {n}_{1}) = \textbf{1}_{({t}_{1},l({n}_{1})) \in all\_nodetran (D)} \\[0pt]
f({t}_{1}, {n}_{1}, {t}_{2}, {n}_{2}) = \textbf{1}_{({t}_{1},l({n}_{1}), {t}_{2}, l({n}_{2})) \in all\_edgetran (D)} \\[0pt]
f({t}_{1}, {n}_{1}, {t}_{2}, {n}_{2}, {t}_{3}, {n}_{3}) = \\ \textbf{1}_{({t}_{1}, l({n}_{1}), {t}_{2}, l({n}_{2}), {t}_{3}, l({n}_{3})) \in all\_triangletran (D)} \nonumber
\end{gathered}
\end{equation}
\normalsize
\noindent where $t_i$ corresponds to the repair transform associated with node $n_i$. For edge clique, $n_1$ and $n_2$ are parent and child node respectively. For triangle clique, $n_1$, $n_2$ and $n_3$ are parent, left child and right child node respectively. Note that in the remaining of this paper, we use the same notation as here.

The above defined indicator-based feature functions do not take the value of the nodes into account. To study the repair transforms associated with triangle clique when the value of the left child is the same with that of the right child (e.g., same variable access), we define another set of  indicator-based feature functions for triangle clique as follows:
\begin{equation}
\begin{split}
{triangletran}^{spe} (\textbf{t}, \textbf{n}) = \{(T_1, L_1,T_2,L_2,T_3, L_3) | \exists \emph{n} \in \textbf{n}, \emph{i} \in \\ \mathbb N:   l(n)=L_1, t(n)=T_1, l(n^{\mapsto i})=L_2, t(n^{\mapsto i})=T_2, l(n^{\mapsto (i+1)}\\) =L_3, t(n^{\mapsto (i+1)})=T_3, L_2=L_3, v(n^{\mapsto i}) = v(n^{\mapsto (i+1)})\}  \\
& 
\nonumber
\end{split}
\end{equation}
\vspace{-7mm}
\begin{equation}
\begin{split}
{all\_triangletran}^{spe} (D) = \cup_{i=1}^m {triangletran}^{spe} (\textbf{t}^i, \textbf{n}^i) \\
& 
\nonumber
\end{split}
\end{equation}
\vspace{-7mm}
\begin{equation}
\begin{split}
{f}^{spe}({t}_{1}, {n}_{1}, {t}_{2}, {n}_{2}, {t}_{3}, {n}_{3}) = \textbf{1}_{({t}_{1}, l({n}_{1}), {t}_{2}, l({n}_{2}), {t}_{3}, l({n}_{3}))}\\ {}_{\in {all\_triangletran}^{spe} (D) \land  {v}({n}_{2})={v}({n}_{3}) } 
\nonumber
\end{split}
\end{equation}

In the learning phase, we learn the corresponding weight for each of the indicator-based feature functions defined above.
Note that the indicator-based feature functions and the learned weights for them can vary depending on the training data \emph{D}, but they are independent of the buggy code snippet for which we are trying to predict repair transforms. 
\subsection{Learning and Prediction}
\noindent
We in this section describe how to train the CRF model for repair transform prediction and use the already trained CRF model to do prediction. 

\subsubsection{Learning} 
Recall that the learning problem is to determine the optimal weights $\lambda$ = \{$\lambda_{k}\}_{k=1}^K$ for feature functions from the training data $\emph{D}= \{\langle \textbf{t}^{i}, \textbf{n}^{i}\rangle\}_{i=1}^m$. 
The typical way to train CRF model is using classical \emph{penalized maximum (log)-likelihood} \cite{CRF2}, which optimizes the following log-likelihood objective function with respect to the model \emph{p}(\textbf{t}|\textbf{n}, $\lambda$): 
\begin{equation}
\begin{gathered}
l(\lambda)= \sum_{i=1}^m \text{log} p (\textbf{t} = \textbf{t}^{i} | \textbf{n} = \textbf{n}^{i}; \lambda) \nonumber
\end{gathered}
\end{equation}
\normalsize

In other words, the weights for feature functions are chosen such that the training data has the highest probability under the model. 

\textbf{Transform Number Imbalance Issue.} The above objective function treats each $p(\textbf{t}=\textbf{t}^{i}| \textbf{n}=\textbf{n}^{i}; \lambda)$ as equally important. However, one significant characteristic in repair transform prediction problem is that the buggy code snippet is nearly correct, and typically just a few actual repair transforms on certain AST nodes are needed. Note that we attach a virtual `EMPTY' repair transform to those nodes which are not associated with any repair transforms.
Thus, the training data $\emph{D}=\{\langle \textbf{t}^{i}, \textbf{n}^{i}\rangle\}_{i=1}^m$ is skewed in turns of the number of AST nodes that are associated with actual repair transforms. If the skew is too large, the learned weights $\lambda$ = \{$\lambda_{k}\}_{k=1}^K$ will be dominated by those training examples with few repair transforms on few nodes.
However, correctly predicting those instances which need relatively more repair transforms on nodes is important as those bugs are much harder to deal with. We call this issue ``\emph{transform number imbalance issue}''. 

To deal with the training data imbalance problem, there are typically three groups of solutions: sampling methods \cite{sample}, cost-sensitive learning \cite{costlearning}, and one-class learning \cite{oneclass}. For our repair transform prediction problem, we propose a method called "transform distribution aware learning". The method is similar to sampling methods, but does not have the disadvantage of removing important examples in under-sampling and adding redundant examples in over-sampling. The method analyzes the training data before the launch of training and gives more weight on those training examples which have relatively more nodes associated with actual repair transforms. Formally, for the training dataset $\emph{D}= \{\langle \textbf{t}^{i}, \textbf{n}^{i}\rangle\}_{i=1}^m$, we define the set \emph{U} which contains all the observed numbers of actual repair transforms used for repairing a bug as follows:
\begin{equation}
\begin{gathered}
U= \{ n | n \in \mathbb{Z}^+,  \exists \langle \textbf{t}^{i}, \textbf{n}^{i} \rangle \in \emph{D}: S(\textbf{t})=n \}
\nonumber
\end{gathered}
\end{equation}
\normalsize
\noindent where S(\textbf{t}) denotes the number of actual repair transforms in \textbf{t}. 

We then define a distribution-aware prior $\chi_{i}$ as: 
\begin{equation}
\begin{gathered}
\overline{N} = \frac{1}{|U|} \sum_{{u} \in {U}}  N_u \, , \quad 
\chi_{i} = \Big(\frac{\overline{N}}{N_{S(\textbf{t}^i)}}\Big) ^{q}
\nonumber
\end{gathered}
\end{equation}
\normalsize
\noindent where $N_u$ is the number of training examples in \emph{D} that have \emph{u} actual repair transforms, $\overline{N}$ is the average number of training examples per each number in \emph{U}, and \emph{q} is a coefficient that controls
the magnitude of the distribution-aware prior. 

Finally, we multiply the distribution-aware prior with the log probability for each training example $\langle \textbf{t}^{i}, \textbf{n}^{i}\rangle $ in the objective function and get a new objective function:
\begin{equation}
\begin{gathered}
l(\lambda)= \sum_{i=1}^m \chi_{i} \text{log} Pr (\textbf{t} = \textbf{t}^{i} | \textbf{n} = \textbf{n}^{i}; \lambda) \nonumber
\end{gathered}
\end{equation}
\normalsize

Note that when the training dataset \emph{D} has a uniform distribution (i.e., for each ${u} \in {U}$, $N_u$ is equal) or when the coefficient \emph{q} equates to 0, the new objective function is reduced to the typical objective function. 
Through the use of distribution-aware prior, more weights can be put on those training examples which have relatively more actual repair transforms attached to nodes and are scarce in \emph{D}. 
The larger the coefficient \emph{q}, the more weights we put on those types of training examples which are scarce in \emph{D}.
Overall, by using the distribution-aware prior, all the training examples in \emph{D} can be adjusted to have a balanced impact in the learning process. 

\textbf{Regularization.} As our CRF model contains a large number of feature functions, we use \emph{regularization} to penalty on weight vectors whose norm is too large (i.e., avoid 
over-fitting). We use the typical penalty based on the Euclidean norm of $\lambda$ and the strength of the penalty is determined by the parameter 1/2$\delta^2$. The regularized objective function is then:
\begin{equation}
\begin{gathered}
l(\lambda)= \sum_{i=1}^m \sum_{{c} \in {C}} \sum_{k=1}^K \chi_{i} \lambda_{k}f_{k}(\textbf{t}_{c}^{i}, \textbf{n}_{c}^{i}) - 
\sum_{i=1}^m \chi_{i} \text{log} Z(\textbf{n}^{i}) -
\sum_{k=1}^K \frac{\lambda_{k}^2}{2\delta^2}
\nonumber
\end{gathered}
\end{equation}
\normalsize

The above function is concave and every local optimum is also a global optimum. But $l(\lambda)$ in general cannot be maximized in closed form, so numerical optimization is used.  We use the particularly successful method L-BFGS \cite{BFGS}.
L-BFGS belongs to the family of quasi-Newton methods and 
can be used as a black-box optimization routine by feeding the value and the first derivative of the objective function.

The first derivative of the objective function $l(\lambda)$ for each parameter $\lambda_{k}$ is:

\begin{equation}
\begin{gathered}
\frac{\partial l(\lambda)}{\partial \lambda_{k}} = \sum_{i=1}^m \sum_{c \in C} \chi_{i} f_{k}(\textbf{t}_{c}^{i}, \textbf{n}_{c}^{i}) - 
\sum_{i=1}^m \sum_{c\in C} \sum_{\textbf{t}_{c}^{i^+} \in {\Phi_c}} \chi_{i} p (\textbf{t}_{c}^{i^+}| \textbf{n}_{c}^{i} ) \\ \quad \cdot f_{k}(\textbf{t}_{c}^{i^+}, \textbf{n}_{c}^{i})  -
\frac{\lambda_{k}}{\delta^2}
\nonumber
\end{gathered}
\end{equation}

\noindent where ${\Phi_c}$ ranges over all assignments to \textbf{t} in the clique \emph{c}. 
The computation of the first term is straightforward (i.e., sum the feature function values over the training dataset), but calculating the second term requires to calculate the marginal probability $p (\textbf{t}_{c}| \textbf{n}_{c})$, which is an inference task and we will discuss it below.

\subsubsection{Prediction} 
Recall that prediction involves substituting the defined feature functions and the learned weights for them into formula (4) in Section 4.1 to get the conditional probability of each possible transform \textbf{t} for the AST nodes \textbf{n}.

Typically, CRFs output the single most likely prediction by using the MAP (Maximum a Posteriori) \cite{PGM} query: $\textbf{t} = {\text{argmax}_{\textbf{t}' \in {\Omega}_{\textbf{n}}}} \emph{p}(\textbf{t}'|\textbf{n})$.
As said in the section about learning, the learning process needs to calculate the marginal probability $p (\textbf{t}_{c}| \textbf{n}_{c})$ for a certain clique \emph{c}. These are the two inference problems that arise in CRFs, and \emph{can be seen as fundamentally the same operation on two different semirings} \cite{CRF2}. To change the marginalization problem to the maximization problem, we just need to substitute maximization calculation for addition calculation.

When the associated undirected graphs with CRF model have cycles, typically approximate inference algorithms have to be used. However, one advantage of our CRF model is that the maximal clique in the undirected graph is triangle, for which efficient exact inference algorithms are available. The process is first using \emph{junction tree algorithm} to change the graph into a tree, and then \emph{belief propagation algorithm} can be used to do the inference \cite{CRF2}. We refer readers to \cite{junction} and \cite{belief} for details about junction tree algorithm and belief propagation algorithm respectively. 

The belief propagation algorithm is also called message-passing algorithm, and the marginal distributions are recursively computed using messages exchanged between all the nodes in the junction tree. When it comes to the original undirected graphs for CRFs, let \emph{c} be a maximal clique and the set $N^c$ be its neighbour maximal clique set, i.e., the set of maximal cliques that have common nodes with \emph{c}. One can informally interpret message passing as that the marginal distribution of \emph{c} is determined by summing over all the admissible label assignments for the nodes of each $c^, \in N^c$, and for each $c^,$, its marginal distribution in turn relies on all the admissible label assignments for all of its neighbour maximal cliques. 

\textbf{Constraint on Valid Repair Transform.} One important characteristic of the repair transform prediction problem is that the admissible repair transforms assigned to a certain node \emph{n} are highly dependent of the label of itself and the labels of its neighbour nodes. For instance, for our defined 16 repair transforms, a transform \emph{t} on the virtual root node of a statement is valid only when $ t \in $ \{$\tt{Wrap-IF-N}$, $\tt{Wrap-IF-O}$, $\tt{Wrap-IFELSE-N}$, $\tt{Wrap-IFELSE-O}$, $\tt{Wrap-TRY}$ \}. To take this into account, we need to establish constraints to determine whether the joint assignment (\textbf{t}, \textbf{n}) for a certain clique is admissible.  

Constraints restrict the sets of admissible label assignments for cliques and the message passing algorithm can be easily modified to account the constraint: only admissible labeling assignments are considered in the messages. Constraints can be established according to the domain knowledge about what kinds of repair transforms are possible for certain nodes in a clique. 
We in this paper define constraints based on the training dataset $\emph{D}= \{\langle \textbf{t}^{i}, \textbf{n}^{i}\rangle\}_{i=1}^m$. The idea is that the training dataset \emph{D} is representative enough, and for an assignment $\textbf{t}^{k}$ to be admissible, we should have observed the occurrence of this in the training data. Using the notations in Section 5.2.2, for different types of cliques, we define the set of admissible repair transforms for different node labels as follows:
\vspace{-1mm}
\begin{equation}
\begin{split}
 \emph{adm\_{nodetran} (D)}  = \{(L,T^S) | \exists L \in \gamma,  \forall T \in T^S, \\ \textbf{1}_{(T,L)  \in all\_nodetran (D)} \} \\
& 
\nonumber
\end{split}
\end{equation}
\vspace{-7mm}
\begin{equation}
\begin{split}
\emph{adm\_{edgetran} (D)}  = \{(L_1, L_2, T^P) | \exists L_1, L_2 \in \gamma, \forall \langle {T}_{1}, {T}_{2} \rangle \in T^P, \\ \textbf{1}_{({T}_{1},{L}_{1}, {T}_{2}, {L}_{2}) \in all\_edgetran (D)} \} \\
& 
\nonumber
\end{split}
\end{equation}
\vspace{-7mm}
\begin{equation}
\begin{split}
\emph{adm\_{triangletran} (D)}  = \{(L_1,L_2,L_3,T^T) |\exists L_1, L_2,  L_3 \in \gamma, \forall \\\langle {T}_{1}, {T}_{2}, {T}_{3}  \rangle \in T^T,   \textbf{1}_{({T}_{1}, {L}_{1}, {T}_{2}, {L}_{2}, {T}_{3}, {L}_{3}) \in all\_triangletran (D)} \} 
\nonumber
\end{split}
\end{equation}
where $\gamma$ represents the set of possible labels for a node, $T^S$, $T^P$, and $T^T$ are the sets whose elements are 1-tuple, 2-tuple, and 3-tuple respectively. 

Based on the established admissible repair transforms for different node labels, we can then determine whether the assignments for different cliques violate the constraint as follows:

\vspace{-3mm}
\begin{equation}
\begin{split}
 V({t}_{1}, {n}_{1})  = \textbf{1}_{\exists (L,T^S) \in adm\_{nodetran} (D): l({n}_{1})=L \land {t}_{1} \notin T^S}\\
& 
\nonumber
\end{split}
\end{equation}
\vspace{-7mm}
\begin{equation}
\begin{split}
V({t}_{1}, {n}_{1}, {t}_{2}, {n}_{2}) = \textbf{1}_{\exists (L_1, L_2, T^P) \in adm\_{edgetran} (D): l({n}_{1})=L_1 \land }\\{}_{l({n}_{2})=L_2 \land ({t}_{1}, {t}_{2}) \notin T^P}  \\
& 
\nonumber
\end{split}
\end{equation}
\vspace{-7mm}
\begin{equation}
\begin{split}
V({t}_{1}, {n}_{1}, {t}_{2}, {n}_{2}, {t}_{3}, {n}_{3}) = \textbf{1}_{\exists (L_1,L_2,L_3,T^T) \in adm\_{triangletran} (D): }\\{}_{l({n}_{1})=L_1 \land l({n}_{2})=L_2 \land l({n}_{3})=L_3 \land ({t}_{1}, {t}_{2}, {t}_{3}) \notin T^T}
\nonumber
\end{split}
\end{equation}

\noindent where the value 1 means the constraint is violated and the assignment is not admissible. 

The use of constraint arises for two reasons. First, it can significantly reduce the time complexity in inference as only admissible transform assignments are considered in the messages. Second, it allows to eliminate incorrect transform assignments according to domain knowledge, resulting in accuracy improvement. 

\section{Experimental Evaluation}
\noindent
We present in this section the implementation details, the experimental methodology, and the evaluation results.

\subsection{Implementation.} 
\noindent
We implemented our repair transform prediction approach in a tool called $\tt{Seer}$. The tool is written in Java and currently works for Java code. It consists of two parts: repair transform extraction and CRF model learning and prediction. For repair transform extraction, we use GumTree \cite{gumtree} to extract the AST tree edit script. GumTree is an off-the-shelf, state-of-the-art tree differencing tool that computes AST-level program modifications, and outputs them as the 4 basic tree edit operations: UPD, ADD, DEL, and MOV. We also use 
Spoon \cite{spoon} to analyze the code that surrounds the AST nodes affected by tree edit operations. Besides, PPD \cite{madeiral2018towards} is employed to facilitate the detection of certain repair transforms. Our CRF model is implemented on top of the XCRF library \cite{xcrf}, which is a framework for building CRFs to label XML data. We extend XCRF library to incorporate our specific feature functions, learning algorithm and prediction algorithm. 
In particular, a major modification both at the conceptual and implementation level is the support for computing the \emph{top-k} predictions, i.e., the predictions with the \emph{k} highest conditional probabilities. 

\subsection{Experimental Setup.}

\noindent
\textbf{Dataset.} {We use the Boa dataset \cite{boa,megadiff} as the source to get the required dataset with repair transforms on AST nodes. Boa is a domain-specific language and infrastructure for analyzing ultra-large-scale software repositories and the Boa dataset includes 4,590,679 bug fixing commits. To compute the diffs of the bug-fixing commits with GumTree, we separately compute the diff between each file affected by a commit (i.e., a patched file) and its previous version (i.e., a buggy file). The output of GumTree is an \emph{edit script} composed by \emph{tree edit operations}.
We find that when the diff is relatively large, the edit scripts are frequently not accurate enough to reflect real code changes. In addition, a bug-fixing commit with large diffs is much more likely to include irrelevant changes such as feature additions or refactorings~\cite{just2014defects4j}.} Therefore, we limit our repair transform extraction to those bug-fixing commits with relatively small diffs. We experimented with different thresholds for root tree edit operations and find when the threshold is set to 10, the GumTree outputs are accurate enough to reflect real code changes in most cases and we can correctly attach repair transforms to AST nodes. After setting the threshold to be 10, we find that the tree edit operations in a certain file are typically targeting a single statement. Consequently, we use the AST of the targeted statement as the AST of the buggy code. In case occasionally the tree edit operations in a file involve multiple statements, we view each of these statements as an isolated bug. Finally, we have 267,555 pairs of $\langle \textbf{t}, \textbf{n} \rangle$ extracted from the original bug fixing commits, where \textbf{n} is the AST of a changed statement and \textbf{t} is the set of repair transforms associated with the nodes of \textbf{n}. Note that our approach itself is applicable to ASTs of any code snippet.

\autoref{tab:summary} shows the number for different repair transforms we have in the dataset. The "Single" (resp. "Multiple") refers to the case where the number of actual repair transforms for a certain $\langle \textbf{t}, \textbf{n} \rangle$ is one (resp. larger than one).  Note that a single repair transform may involve multiple root tree edit operations according to our definitions in \autoref{tab:definitions-transform-inner} and \autoref{tab:definitions-transform-virtual}, thus the "Single" case does not necessarily mean that the corresponding bug-fixing commit just contains one root tree edit operation. We can see from \autoref{tab:summary} that a majority of the data involves just one actual repair transform applied to a certain node. To check the quality of the dataset, we randomly sample 200 examples for each repair transform and also 200 examples for "multiple-transform" case, and manually check whether correct repair transforms are attached to correct AST nodes. {More specifically, for the AST \textbf{n} of a changed statement, we carefully examine the bug-fixing commit and determine what kinds of repair transforms are associated with the nodes of \textbf{n}, i.e., we manually establish the ground-truth repair transforms $\textbf{t}^{real}$ associated with \textbf{n}. We compare the automatically extracted $\langle \textbf{t}, \textbf{n} \rangle$ with the manually established $\langle \textbf{t}^{real}, \textbf{n} \rangle$ and if they are exactly the same (i.e., each AST node is associated with exactly the same repair transform, including EMPTY which means no repair transform), we view correct repair transforms have been attached to the correct AST nodes. The result of manual check is shown in the column \emph{\#Correct}. Overall, we can see that the accuracy is satisfactory, achieving 88\% correctness rate for "multiple-transform" case and at least 91\% correctness rate for "single-transform" case.}

\begin{table*}
\centering
 \caption{Descriptive statistics of the repair transforms in the dataset.}
   \vspace{-0.6\baselineskip}
  \label{tab:summary}
  \small
    \setlength\extrarowheight{-4pt}
  \begin{tabular}{|l|c|c||l|c|c|}
    \hline
    Repair Transform & \#Number & \#Correct & Repair Transform & \#Number & \#Correct \\
    \hline
    $\tt{Wrap-IFELSE-N}$ & 4,528 & 90\% & $\tt{LogExp-Exp}$ & 21,331& 98\% \\
    $\tt{Wrap-IFELSE-O}$ & 11,366 & 92\% & $\tt{LogExp-Red}$& 3,237& 97\% \\
    $\tt{Unwrap-Meth}$ & 1,658& 92\% & $\tt{Wrap-IF-N}$ & 10,307 & 93\% \\
    $\tt{Constant-Rep}$ & 84,248& 96\%  & $\tt{Wrap-IF-O}$ & 11,653& 95\% \\
    $\tt{Meth-RW-Var}$ & 4,084& 91\% & $\tt{Wrap-Meth}$ & 6,403 & 95\% \\
    $\tt{Meth-RW-Meth}$ & 43,314& 92\% & $\tt{Var-RW-Var}$ & 29,322 & 96\% \\
    $\tt{BinOper-Rep}$ & 5,928& 97\% & $\tt{Unwrap-IF}$ & 6,936& 96\% \\
    $\tt{Var-RW-Meth}$ & 3,694 & 93\% & $\tt{Wrap-TRY}$ & 6,662 & 96\% \\
    \hline
    \hline
    $\tt{Single}$ & 254,671 & --- & $\tt{Multiple}$ & 12,884& 88\% \\
    \hline
\end{tabular}
\vspace{-3mm}
\end{table*}

\vspace{1mm}
\noindent
\textbf{Cross-validation.}
We use cross-validation to select parameters of the CRF model and evaluate the performance of the trained CRF model. We split the whole dataset into 10 equal folds, with each fold having the same number of "single-transform" instances for each kind of the 16 considered repair transforms and same number of "multiple-transform" instances. We select one fold as test dataset to see the model performance. 

The other 9 folds in the dataset are used as the training dataset and we use cross-validation to investigate the impact of the parameters and select them accordingly. There are three parameters involved in our CRF model: the regularization parameter $\delta^2$, the parameter \emph{q} that indicates the magnitude of the distribution-aware prior, and finally the L-BFGS iteration parameter \emph{G} that specifies the number of gradient computations used by the optimization procedure. Higher values of $\delta^2$ will incur larger penalty on large weights associated with feature functions and higher values of \emph{G} will result in higher execution cost for inference. We evaluate the error rate on each fold by training a model on the other 8 folds. The error rate is established using the \emph{top-3} evaluation metric described later. We repeat this process by using a set of different parameters and aim to identify parameters with the lowest error rate. The procedure determines that 500 and 200 are good values for $\delta^2$ and \emph{G} respectively, and the following results are based on these two values. The parameter \emph{q} significantly impacts the prediction result for "single-transform" and "multiple-transform" instances, and we will discuss its impact in detail in the result section. 

\vspace{1mm}
\noindent
\textbf{Evaluation Metric.}
For each tuple $\langle \textbf{t}, \textbf{n} \rangle$ in the test dataset, we view a prediction $\langle \textbf{t}', \textbf{n} \rangle$ by our trained CRF model as correct if $\textbf{t} = \textbf{t}'$. 
That is, for each node $ n \in \textbf{n}$, the predicted repair transform associated to it is exactly the same as the ground truth repair transform extracted from data. 
We use both metrics \emph{top-1} and \emph{top-3} to evaluate the model performance. For metric \emph{top-1}, the prediction is considered as a success only if the top 1 prediction (i.e., the prediction with the highest probability) is correct. For metric \emph{top-3}, we consider the prediction as a success if at least one of the top 3 predictions (i.e., the three predictions with the three highest probabilities) is correct. We respectively calculate the number of tuples in the test dataset that have been successfully predicted for these two metrics, and then respectively divide the two numbers by the total number of tuples in the test dataset to see model accuracy. These two metrics have been widely used to evaluate model performance, in particular for hard learning task such as ours \cite{deepdelta, popl16martin,icml16convolutional}.

\vspace{1mm}
\noindent
\textbf{Baseline.}
\emph{History probability baseline:} We first establish a baseline which assigns repair transforms to nodes according to basic statistics on past commits. 
Using the same dataset employed for training the model, we establish a set of tuples <\emph{L}, \emph{T}, \emph{P}> where \emph{L} is a certain node label, \emph{T} is a certain repair transform, and \emph{P} represents the probability of assigning \emph{T} to nodes with label \emph{L}. Basically, let $\emph{N}^{0}$ denote the number of nodes with label \emph{L} in the training dataset and $\emph{N}^{1}$ denote the number of nodes with label \emph{L} and are associated with transform \emph{T}, then \emph{P} = $\emph{N}^{1}$/$\emph{N}^{0}$. The baseline first assumes there needs only one repair transform to correct the program and prioritizes the node \emph{n} and repair transform \emph{T} on it according to the established probability for <\emph{l}(\emph{n}), \emph{T}>. When there are several nodes of the same label, it breaks ties according to their orders during pre-order traversal. After all possible single node transforms have been explored, 
the baseline gradually explores all possible two node transforms and so on. 

{\emph{NMT baseline:} We also build another baseline based on neural machine translation (NMT)~\cite{tufano2019empirical}, in particular the sequence to sequence (Seq2Seq) translation model~\cite{deepdelta}. } To establish this baseline, we first use pre-order to traverse the AST for a certain code snippet and get the linearized AST as a sequence. Then, we use one encoder for the AST node labels and one encoder for the AST node values. The two encoder outputs are combined and fed into the decoder which outputs a sequence of code transforms applied to each AST node accordingly. To train and evaluate the NMT baseline, we use the same training and test dataset as we used to build the CRF model. The NMT baseline is built on top of OpenNMT \cite{opennmt}.

\vspace{1mm}
\noindent
\textbf{Partial Model and Partial Data Comparison.} To see the impact of feature functions, we also compare our "Full Model" against a "Partial Model" that uses only observation-based or indicator-based feature functions. Besides, we also study the impact of training data size on model performance by keeping test data the same but only using 30\% and 70\% of the training data to build the model. 

\subsection{Experimental Results}
\label{sec:eval-results}
\noindent
Our training is run on a cluster node with 24 cores and 512 GB RAM, and the system is Ubuntu 16.04. For the full model, the training takes around 41 hours for each value of parameter \emph{q} when 500 and 200 are set as the values for $\delta^2$ and \emph{G} respectively, and the average prediction time is 48ms.

{\autoref{tab:top1} and \autoref{tab:top3} respectively show the performance of the model when \emph{top-1} and \emph{top-3} are used as the evaluation metric.} The numbers in the cells of the tables represent the prediction accuracy for each kind of transform in "single-transform" category and also the accuracy for "multiple-transform" category (at the table bottom). The columns "History" and "NMT" represent the results obtained by history probability baseline and NMT baseline respectively. The "CRF Full Model" refers to the model that uses both observation-based and indicator-based feature functions, and regardless of the used evaluation metric, the overall optimal model is obtained when the distribution-aware prior coefficient \emph{q} is set to be 0.5 (highlighted with gray). The "Partial Model" refers to the model that uses only observation-based feature functions (denoted as `O') or only uses indicator-based feature functions (denoted as `I'), and the "Partial Data" refers to the model performance when a subset of the original training data is used to build the model. For space reason, we only show the result obtained with the overall optimal value of \emph{q} (that is 0.5) for partial models and models obtained with partial training data. 

\begin{table*}
\caption{The accuracy for AST-level repair transform prediction using Top-1 evaluation metric}
  \label{tab:top1}
   \vspace{-0.6\baselineskip}
\small
\begin{center}
  \setlength\extrarowheight{-4pt}
\begin{tabular}{|l|c|c|c|>{\columncolor[gray]{0.8}}c|c|p{6mm}|p{6mm}|p{4.5mm}|p{4.5mm}|}
\hline
\multirow{3}{*}{Repair Transform}
&\multicolumn{9}{c|}{\bf Transform Prediction Result}\\
\cline{2-10}
&\multicolumn{2}{c|}{Baselines} 
&\multicolumn{3}{c|}{CRF Full Model} 
&\multicolumn{2}{c|}{Partial Model}
&\multicolumn{2}{c|}{Partial Data} 
\\
\cline{2-10}
&History&NMT&\emph{q}=0.0&\emph{q}=0.5&\emph{q}=1.0 & O & I & 30\% & 70\% \\ \hline
 $\tt{Wrap-Meth}$ & 0.0 & 0.06 & 0.06 & 0.06 & 0.03 & 0.01 & 0.03 & 0.05 & 0.06 \\
    $\tt{Unwrap-Meth}$ & 0.0 & 0.07 &0.08 & 0.08 & 0.06 & 0.03 & 0.01 & 0.07 &  0.07 \\
    $\tt{Var-RW-Var}$ & 0.09 & 0.29 & 0.31 & 0.30 & 0.27 & 0.13 & 0.27 & 0.28  & 0.29\\
    $\tt{Var-RW-Meth}$ & 0.0 & 0.10 & 0.13 & 0.15 & 0.11 & 0.00 & 0.16 & 0.15 & 0.14\\
    $\tt{Meth-RW-Var}$ & 0.0 & 0.06 & 0.06 & 0.06 & 0.03 & 0.02 & 0.04 & 0.04  & 0.05\\
    $\tt{Meth-RW-Meth}$ & 0.52 & 0.48 & 0.57 & 0.58 & 0.53 & 0.25 & 0.53 & 0.54 & 0.55 \\
    $\tt{BinOper-Rep}$ & 0.01 & 0.22 & 0.24 & 0.23 & 0.15 & 0.01 & 0.09 & 0.22  & 0.22\\
    $\tt{Constant-Rep}$ & 0.77 & 0.68 & 0.76 & 0.76 & 0.67 & 0.51 & 0.75 & 0.75 & 0.75\\
    $\tt{LogExp-Exp}$ & 0.47 & 0.59 & 0.93 & 0.92 & 0.72 & 0.38 & 0.91 & 0.89  & 0.90 \\
    $\tt{LogExp-Red}$ & 0.0 & 0.09 & 0.10 & 0.10 & 0.11 & 0.02 & 0.08 & 0.08 & 0.09\\
    $\tt{Wrap-IF-N}$ & 0.0 & 0.26 & 0.28 & 0.27 & 0.12 & 0.13 & 0.09 & 0.23 & 0.25\\
    $\tt{Wrap-IF-O}$ & 0.05 & 0.14 & 0.15 & 0.15 & 0.08 & 0.11 & 0.09 & 0.13  & 0.15\\
     $\tt{Wrap-IFELSE-N}$ & 0.0 & 0.05 & 0.05 & 0.05 & 0.03 & 0.02 & 0.00 & 0.04  & 0.05 \\
    $\tt{Wrap-IFELSE-O}$ & 0.0 & 0.04 & 0.05 & 0.05 & 0.02 & 0.02 & 0.01 & 0.03  & 0.04\\
    $\tt{Unwrap-IF}$ & 0.01 & 0.18 & 0.21 & 0.20 & 0.13 & 0.83 & 0.18 & 0.16  & 0.19\\
    $\tt{Wrap-TRY}$ & 0.0& 0.11 & 0.13 & 0.12 & 0.05 & 0.04 & 0.00 & 0.10  & 0.11  \\
    \hline
    \tt{Average} & 0.12 & 0.21 & 0.26 & 0.26 & 0.19 & 0.15 & 0.20 & 0.23 & 0.24 \\
    \hline
    \hline
    $\tt{Mul-Transform}$ & 0.0 & 0.15 & 0.11 & 0.19 & 0.28 & 0.04 & 0.15 & 0.16 & 0.17 \\
    \hline
\end{tabular}
\end{center}
\end{table*}

\begin{table*}
\caption{The accuracy for AST-level repair transform prediction using top-3 evaluation metric}
  \label{tab:top3}
   \vspace{-0.6\baselineskip}
\small
\begin{center}
  \setlength\extrarowheight{-4pt}
\begin{tabular}{|l|c|c|c|>{\columncolor[gray]{0.8}}c|c|p{6mm}|p{6mm}|p{4.5mm}|p{4.5mm}|}
\hline
\multirow{3}{*}{Repair Transform}
&\multicolumn{9}{c|}{\bf Transform Prediction Result}\\
\cline{2-10}
&\multicolumn{2}{c|}{Baselines} 
&\multicolumn{3}{c|}{CRF Full Model} 
&\multicolumn{2}{c|}{Partial Model}
&\multicolumn{2}{c|}{Partial Data} 
\\
\cline{2-10}
&History&NMT&\emph{q}=0.0&\emph{q}=0.5&\emph{q}=1.0 & O & I & 30\% & 70\% \\ \hline
 $\tt{Wrap-Meth}$ & 0.0 & 0.24 & 0.29 & 0.28 & 0.23 & 0.06 & 0.23 & 0.23  & 0.26 \\
    $\tt{Unwrap-Meth}$ & 0.0 & 0.15 &0.23 & 0.26 & 0.17 & 0.05 & 0.04 & 0.16 &  0.21 \\
    $\tt{Var-RW-Var}$ & 0.32 & 0.52 & 0.58 & 0.60 & 0.57 & 0.31 & 0.50 & 0.53  & 0.56\\
    $\tt{Var-RW-Meth}$ & 0.0 & 0.30 & 0.41 & 0.44 & 0.38 & 0.15 & 0.34 & 0.36  & 0.41\\
    $\tt{Meth-RW-Var}$ & 0.0 & 0.26 & 0.30 & 0.30 & 0.23 & 0.04 & 0.20 & 0.24  & 0.27\\
    $\tt{Meth-RW-Meth}$ & 0.89 & 0.77 & 0.84 & 0.83 & 0.79 & 0.64 & 0.75 & 0.79 & 0.80 \\
    $\tt{BinOper-Rep}$ & 0.14 & 0.55 & 0.65 & 0.65 & 0.59 & 0.05 & 0.60 & 0.61  & 0.63\\
    $\tt{Constant-Rep}$ & 0.93 & 0.88 & 0.90 & 0.89 & 0.87 & 0.78 & 0.80 & 0.86 & 0.88\\
    $\tt{LogExp-Exp}$ & 0.96 & 0.89 & 0.98 & 0.98 & 0.91 & 0.78 & 0.93 & 0.92  & 0.95 \\
    $\tt{LogExp-Red}$ & 0.0 & 0.74 & 0.85 & 0.81 & 0.58 & 0.05 & 0.70 & 0.71 & 0.76\\
    $\tt{Wrap-IF-N}$ & 0.19 & 0.52 & 0.59 & 0.56 & 0.39 & 0.43 & 0.31 & 0.50 & 0.55\\
    $\tt{Wrap-IF-O}$ & 0.59 & 0.48 & 0.58 & 0.54 & 0.32 & 0.46 & 0.42 & 0.46  & 0.52\\
     $\tt{Wrap-IFELSE-N}$ & 0.0 & 0.09 & 0.14 & 0.14 & 0.08 & 0.07 & 0.05 & 0.09  & 0.12 \\
    $\tt{Wrap-IFELSE-O}$ & 0.01 & 0.19 & 0.21 & 0.19 & 0.13 & 0.10 & 0.04 & 0.16  & 0.19\\
    $\tt{Unwrap-IF}$ & 0.93 & 0.75 & 0.85 & 0.81 & 0.54 & 0.93 & 0.68 & 0.79  & 0.82\\
    $\tt{Wrap-TRY}$ & 0.0& 0.24 & 0.26 & 0.26 & 0.16 & 0.13 & 0.01 & 0.22  & 0.24  \\
    \hline
    \tt{Average} & 0.31 & 0.47 & 0.54 & 0.53 & 0.43 & 0.32 & 0.41 & 0.47  & 0.51 \\
    \hline
    \hline
    $\tt{Mul-Transform}$ & 0.0 & 0.36 & 0.21 & 0.41 & 0.48 & 0.16 & 0.33 & 0.37 & 0.39 \\
    \hline
\end{tabular}
\end{center}
\end{table*}

\vspace{1mm}

\noindent
\textbf{Comparison with Baselines.} First, we can see that our optimal model overall performs much better than the history probability baseline. The history probability baseline provides some reasonable prediction accuracy only for a few repair transforms (including $\tt{LogExp-Exp}$, $\tt{Constant-Rep}$, $\tt{Meth-RW-Meth}$, $\tt{Unwrap-IF}$, and $\tt{Wrap-IF-O}$) that are most widely used, and performs poorly for "multiple-transform" prediction as it needs to firstly explore all possible single node transforms. Our model, however, not only keeps high prediction accuracy for those few widely used transforms, but also provides good prediction accuracy for those comparably less used transforms and "multiple-transform" instances.

Second, we can see that our optimal model performs consistently better than the NMT baseline for each of the 16 considered repair transforms. When \emph{top-1} is used as the evaluation metric, the NMT baseline on average achieves 21\% and 15\% prediction accuracy for "single-transform" and "multiple-transform" instances respectively, our optimal model instead averagely achieves 26\% and 19\% prediction accuracy for "single-transform" and "multiple-transform" instances respectively. When \emph{top-3} is used as the evaluation metric, the NMT baseline on average achieves 47\% and 36\% prediction accuracy for "single-transform" and "multiple-transform" instances respectively, while our optimal model achieves on average 53\% and 41\% prediction accuracy for "single-transform" and "multiple-transform" instances respectively. Recall that compared with NMT baseline which takes a flattened sequence of tokens as inputs, the main difference of $\tt{Seer}$ is to explicitly take the code structure into account. Since $\tt{Seer}$ performs better than the NMT baseline, this shows that taking the code structure into consideration is beneficial for predicting AST-level code transform.

\vspace{1mm}
\noindent
\textbf{Impact of Distribution-aware Prior.} The distribution-aware prior is introduced to make "single-transform" and "multiple-transform" training examples have a balanced impact in the learning process. The results in both  \autoref{tab:top1} and \autoref{tab:top3} show that the use of distribution-aware prior indeed improves the model performance on "multiple-transform" instances. For instance, by using the magnitude \emph{q} = 1.0, {the model performance for "multiple-transform" instances has gone from 11\% (when \emph{q} = 0.0) to 28\% when \emph{top-1} is used as the evaluation metric} and has gone from 21\% (when \emph{q} = 0.0) to 48\% when \emph{top-3} is used as the evaluation metric. Note, however, that when the magnitude of \emph{q} is too large, it brings an obvious performance degradation for "single-transform" instances. In this case, the "multiple-transform" instances instead have become the dominating class in the training data. Considering the accuracy for both "single-transform" and "multiple-transform" instances, \emph{setting the magnitude q to 0.5 brings overall good performance on our dataset}.

\vspace{1mm}
\noindent
\textbf{Impact of Feature Function.} 
Comparing the performance of the full model with that of the partial model, we can see that the performance has obviously degraded when only one kind of feature functions is used. Overall, the observation-based and indicator-based feature functions respectively perform well for certain repair transforms, and complement each other to form a strong model. In particular, the partial model which only uses observation-based feature functions performs better for repair transforms that target the whole statement (e.g. repair transforms  $\tt{Wrap-TRY}$ and $\tt{Unwrap-IF}$) than the partial model which only uses indicator-based feature functions, while the latter instead performs better for other repair transforms. To sum up, \emph{considering both  observation-based and indicator-based feature functions in conjunction is essential to get good performance.}

\vspace{1mm}
\noindent
\textbf{Impact of Training Data Size.} 
Comparing the performance of the model trained using the whole data with that of the model trained using partial data, we can see that the prediction accuracy of the model has improved when more training data are used.
{When 100\%, 70\%, and 30\% of the whole training dataset is used to train the model and \emph{top-1} is used as the evaluation metric, the established CRF model on average achieves 26\%, 24\%, and 23\% prediction accuracy for 
"single-transform" instances respectively, and averagely achieves 19\%, 17\%, and 16\% prediction accuracy for "multiple-transform" instances respectively. }
When 100\%, 70\%, and 30\% of the whole training dataset is used to train the model and \emph{top-3} is used as the evaluation metric, the established CRF model on average achieves 53\%, 51\%, and 47\% prediction accuracy for 
"single-transform" instances respectively, and averagely achieves 41\%, 39\%, and 37\% prediction accuracy for "multiple-transform" instances respectively. Overall, 
\emph{the amount of training data is critical to model performance.}

\vspace{1mm}
\noindent
\textbf{Summary.} In summary, this set of large scale experiments has shown that
1) our model performs well for vastly different repair transforms, and and works well for both "single-transform" and "multiple-transform" code change scenarios; 
2) our model performs consistently better than all baselines, including the strong NMT baseline, suggesting the importance of considering code structure in achieving good prediction accuracy for code transform; 
3) the distribution-aware prior coefficient \emph{q} is an important configuration parameter in the model, and our experiments show that $q=0.5$ is a good value; 
4) the use of both observation-based and indicator-based feature functions, i.e. the mix of automatically extracted structural features and carefully engineered code analysis features, is key to obtain good performance; 
5) the size of training data significantly impacts model performance and more training data will be beneficial.

\subsection{Case Studies}
\begin{figure*}
\centering
\makebox[0pt]{ \includegraphics[scale=0.245]{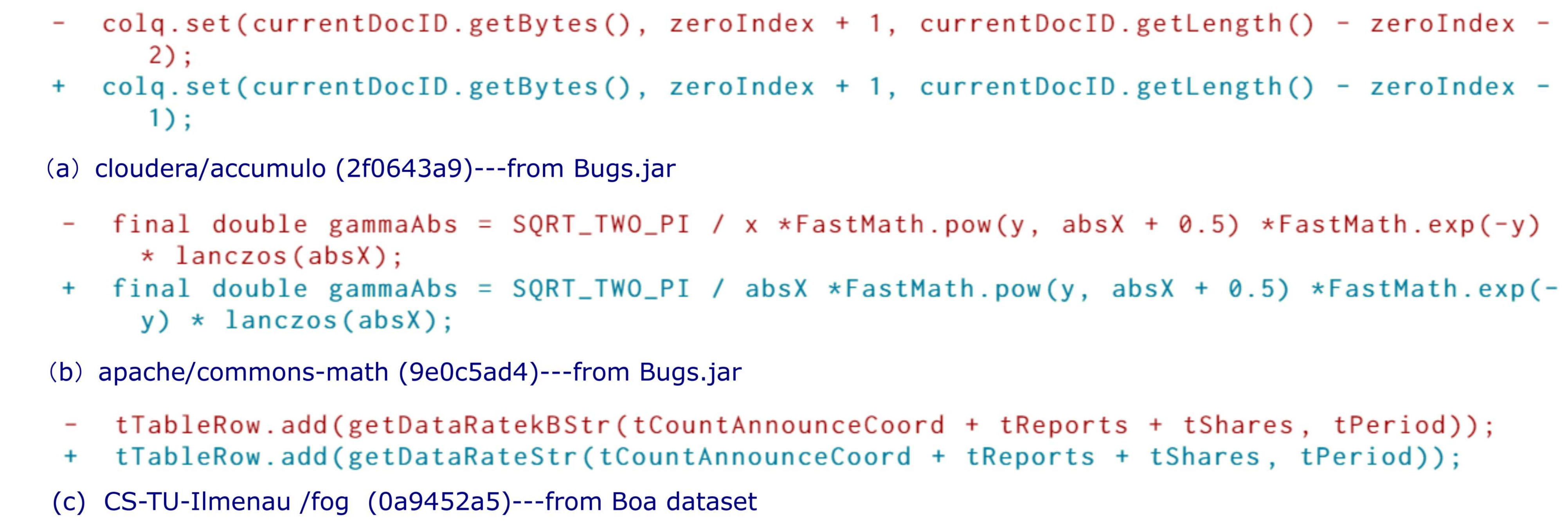}}
 \caption{
  Examples to illustrate the ability of the  CRF model in predicting "single-transform" instances. 
}
 \label{fig:singletran}
\end{figure*}

\noindent
We now build on seven examples to illustrate the performance of our established CRF model. Besides the Boa dataset, we also use examples from Defects4J~\cite{just2014defects4j} and Bugs.jar~\cite{Elixir}, two widely used benchmarks in the program repair community. For all these examples, the most likely prediction by our overall best performance CRF model (\emph{q} = 0.5) contains exactly the ground-truth repair transform(s) needed for correcting each bug. In other words, our best performance CRF model makes perfect prediction for all these seven examples. With regard to the NMT baseline, the top 3 predictions by it for each of these seven examples do not contain the ground-truth repair transform(s) needed for correcting the bug. For history probability baseline, it places the ground-truth repair transform needed for correcting the bug example in \autoref{fig:singletran} (a) or \autoref{fig:singletran} (c) as the second most likely prediction, and the top 3 predictions by it for each of the other five examples do not contain the ground-truth repair transform(s) needed for correcting the bug. 

\noindent
\textbf{Where to Apply Transform and What Transform to Apply?} {The three examples in \autoref{fig:singletran} (a), \autoref{fig:singletran} (b), and \autoref{fig:singletran} (c) respectively require a single repair transform to a constant, variable access, and method call. There are a wide variety of program elements in the input code snippets for all these three examples that can possibly be subject to our defined repair transforms (i.e., the defined repair transforms can possibly apply to multiple program elements), and in particular there are respectively more than one constant, variable access, and method call in the input code snippets for \autoref{fig:singletran} (a), \autoref{fig:singletran} (b), and \autoref{fig:singletran} (c), but our established CRF model can accurately predict which specific program element should be  subject to repair transforms. In addition, the program elements constant, variable access, and method call all can be subject to various repair transforms defined in this paper, our established CRF model can precisely predict the specific repair transforms needed. Overall, our established CRF model is capable of exactly predicting both "where to apply code transform" and "what code transform to apply".}

\begin{figure*}
\centering
\makebox[0pt]{ \includegraphics[scale=0.245]{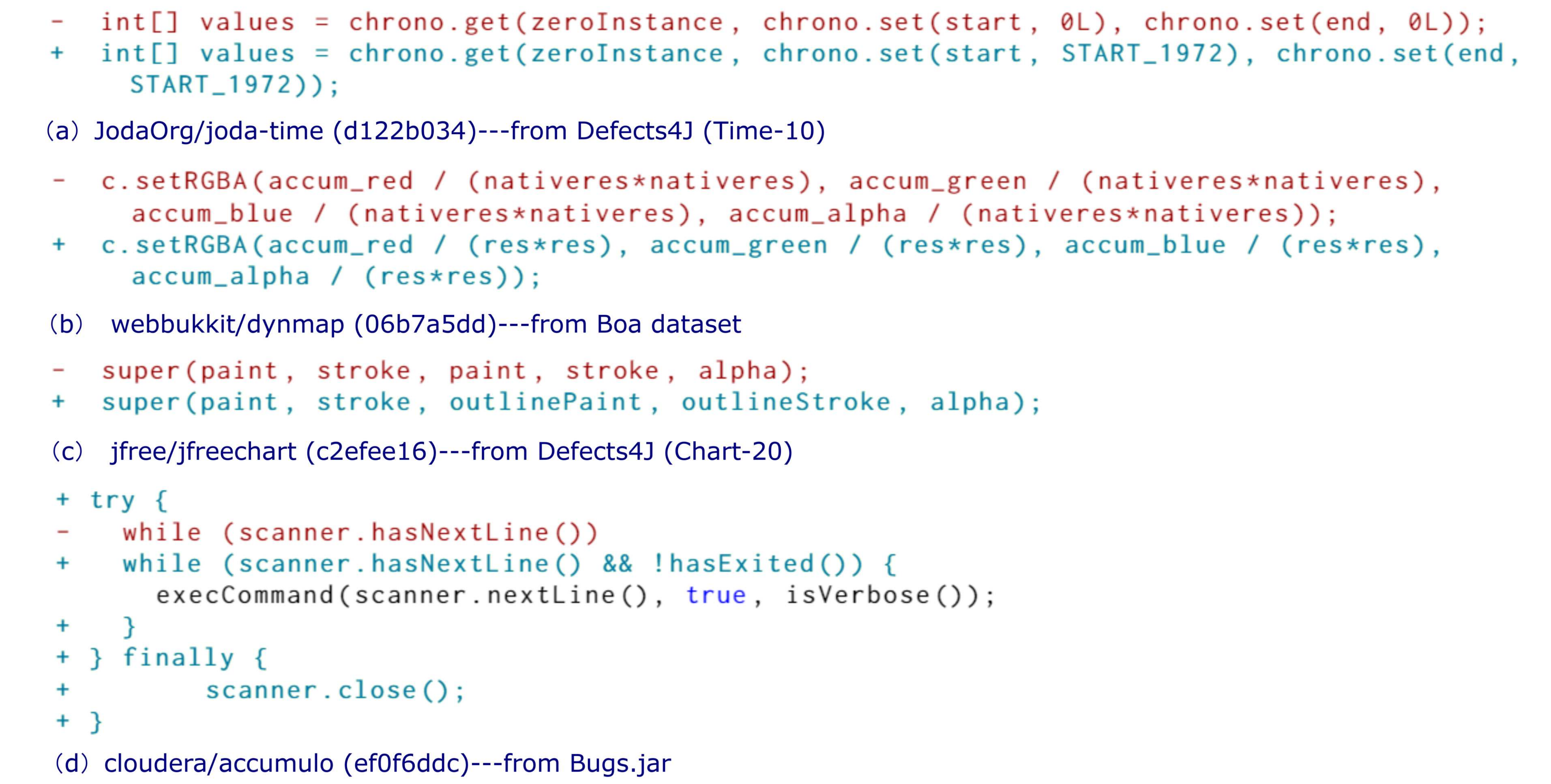}}
 \caption{
   Examples to illustrate the ability of the CRF model in predicting "multiple-transform" instances.  
}
 \label{fig:multipletran}
\end{figure*}

\noindent
\textbf{Co-Transform} {The four examples in \autoref{fig:multipletran}
all require multiple repair transforms to be simultaneously applied. For the two examples in \autoref{fig:multipletran} (a) and \autoref{fig:multipletran} (b), there is a need for multiple repair transforms to instances of the same program element and our established CRF model can accurately predict this kind of co-transform. In particular, note that there is a need for 8 $\tt{Var-RW-Var}$ transforms to the variable access "nativeres" in \autoref{fig:multipletran} (b), our established CRF model is able to establish this fact. For the example in \autoref{fig:multipletran} (c), the input code snippet contains multiple instances of the same program element (including variable access "paint" and variable access "stroke") and it requires repair transforms to some specific but not all instances of a certain program element: our established CRF model is capable of precisely predicting this kind of co-transform. For the example in \autoref{fig:multipletran} (d), multiple radically different types of repair transforms are needed to be applied to different kinds of program elements: our established CRF model can also accurately predict this kind of co-transform. In short, our established CRF model can recognize the necessity of applying multiple repair transforms to repair a certain bug and can accurately predict different categories of co-transforms.}

\subsection{Usefulness of the Predicted Repair Transforms}
\noindent
To achieve reasonable accuracy and scalability in automatically guiding code changes, we in this paper abstract low-level AST edits into high-level code transforms and separate code change generation into high-level code transform prediction and high-level code transform concretization phases. The accurate prediction of high-level code transform is the essential prerequisite for the concretization phase, and is the major focus of our work in this paper. We propose a structured prediction approach based on CRF in order to accurately predict the high-level code transforms, and the approach specifically takes two types of code features into account. Our systematic evaluation has shown that the established CRF model consistently performs better than several strong baselines for the repair transform case. 
For illustrating the high-level code transform concretization phase and demonstrating the usefulness of the predicted high-level repair transforms, an additional evaluation on the benchmark \emph{Defects4J v1.2} \cite{just2014defects4j} is performed. 

This version of Defects4J consists of 395 bugs from 6 projects (\emph{i.e.}, Chart, Closure, Lang, Math, Time, and Mockito), and has been extensively used for evaluating the effectiveness of various automatic program repair (APR) techniques \cite{Le2016HDRepair, Elixir, wen2018context, sequencer, DLFix, Recoder, CODIT, CoCoNuT}. The specific number of bugs for each of the 6 projects is shown in column "\#Bugs (ori)" of \autoref{tab:dataset}. As our model currently targets the 16 repair transforms described in Section 5.1 and is trained using bug-fixing changes that target a single statement, we manually check the 395 Defects4J bugs to identify ones whose repair changes center around one statement and fall into the targeted 16 repair transforms. Finally, we find that 89 Defects4J bugs satisfy the requirement, and the column "\#Bugs (sel)" of \autoref{tab:dataset} gives the qualified number of bugs for each project. 

For these selected 89 Defects4J bugs, we use the repair transform extraction part of our tool $\tt{Seer}$ to attach repair transforms to AST nodes of the targeted statement, and we also manually verify that correct repair transforms indeed have been attached to correct AST nodes. We then use our optimal model (i.e., full model with $q=0.5$) to predict repair transforms on the AST nodes, and compare the predicted repair transforms on AST nodes with the extracted repair transforms on AST nodes. The two columns "\#(top-1)" and "\#(top-3)" of \autoref{tab:dataset} respectively show the number of bugs that can be successfully predicted by our model for each project when \emph{top-1} and \emph{top-3} are used as the evaluation metrics. Overall, the prediction accuracy for the 89 Defects4J bugs is 42.69\% and 53.93\% respectively for the evaluation metrics \emph{top-1} and \emph{top-3}. The result again shows that our model achieves good accuracy in predicting AST-level code transform by taking code structure into account.

\begin{table}
 \caption{The prediction performance of our optimal model for the selected 89 Defects4J bugs}
  \label{tab:dataset}
  \centering
  \small
  \begin{tabular}{|l|c|c|c|c|r|}
    \hline
    Subjects & \#Bugs (ori) & \#Bugs (sel) & \#(top-1) & \#(top-3) \\
    \hline
    Chart   & 26  & 10 & 4 & 6 \\
    Closure & 133 & 34 & 20 & 23 \\
    Lang    & 65  & 12 & 3 & 4  \\
    Math  & 106  & 21 &  7 & 10  \\
    Time  & 27  & 4 &  1 & 2  \\
    Mockito  & 38  & 8 &  3 & 3  \\
    \hline
    \hline
    Total  & 395  & 89 &  38  & 48 \\
    \hline
\end{tabular}
\end{table}

We then implement a proof-of-concept synthesizer to illustrate the usefulness of the predicted AST-level repair transforms in producing the final patches. Among the 16 targeted repair transforms, we concretize 8 repair transforms to generate the final patch, including $\tt{Unwrap-Meth}$, $\tt{Var-RW-Var}$, $\tt{Var-RW-Meth}$, $\tt{Meth-RW-Meth}$, $\tt{Meth-RW-Var}$,  $\tt{BinOper-Rep}$, $\tt{Constant-Rep}$, and $\tt{LogExp-Red}$. For the other 8 repair transforms, more complex code snippet synthesis algorithms are needed to concretize them into patches and we leave this to future work. 
In particular, the considered 8 repair transforms are concretized as follows:
\begin{itemize}
    \item For repair transform $\tt{Unwrap-Meth}$, we delete the outer method call and keep the expression(s) wrapped inside the method call. If the deleted method call has several different expressions as its arguments, we separately keep each expression in ascending order of the argument index before a plausible patch is found. 
    \item For repair transform $\tt{Var-RW-Var}$, we replace the variable $\emph{v}$ (attached with $\tt{Var-RW-Var}$) with a certain searched variable. The candidate variables are the set of accessible variables at the buggy statement whose types are same as that of $\emph{v}$, and they will be tried in descending order of the similarity with $\emph{v}$. The similarity is established for the identifier names of variables based on the Levenshtein distance.
    \item For repair transform $\tt{Var-RW-Meth}$, we replace the variable $\emph{v}$ (attached with $\tt{Var-RW-Meth}$) with a certain searched method call whose return type is same as the type of $\emph{v}$. Note that whenever we refer to method search, the details about the process will be explained later. 
    \item For repair transform $\tt{Meth-RW-Meth}$, we first replace the method call $m$ (attached with $\tt{Meth-RW-Meth}$) with another searched method call which accepts exactly the same arguments as $m$. The same here specifically refers to same in number and type of arguments. All of the candidate method calls are tried in descending order of the similarity with $m$, and the similarity again is established for the identifier names of method calls based on the Levenshtein distance. If this first procedure does not result in a plausible patch, we then search the file (where the buggy statement resides in) for overloaded methods of $m$ and delete or add certain arguments for $m$ accordingly. These two procedures correspond to the two definitions for repair transform $\tt{Meth-RW-Meth}$ in \autoref{tab:definitions-transform-inner}. 
    \item For repair transform $\tt{Meth-RW-Var}$, we replace the method call $\emph{m}$ (attached with $\tt{Meth-RW-Var}$) with a certain variable whose type is same as the return type of $\emph{m}$. The candidate variables are those that are accessible at the faulty statement, and these variables are tried in ascending order of their occurrence in the file where the faulty statement resides in.
    \item For repair transform $\tt{BinOper-Rep}$, we replace the binary operator $\emph{bo}$ (attached with $\tt{BinOper-Rep}$) with another binary operator. We divide binary operators into 5 kinds: logical (||, \&\&), bit (|, \^, \&), comparison (==, !=, <, >, <=, >=), shift (<<, >>), and arithmetic (+, -, *, /, \%). We only replace $\emph{bo}$ with another binary operator whose kind is the same as that of $\emph{bo}$. 
   \item For repair transform $\tt{Constant-Rep}$, we replace the value of the literal (attached with $\tt{Constant-Rep}$) with another literal value of the same kind. We currently consider three kinds of literal value: numerical, boolean, and string. We only replace literal value of these three kinds, and the specific literal values used for replacing the old literal value are searched in the file where the faulty statement resides in. 
  \item For repair transform $\tt{LogExp-Red}$, we remove the condition(s) of the logical expression until a plausible patch is found. We first try to see whether it is possible to achieve this by removing one condition, then two conditions, and so on up to all conditions. 
\end{itemize}

A number of points deserve comment. First, when we search for methods to be called, the candidate methods include those methods that have been called and those defined but not called methods in the file where the faulty statement resides in. All of the searched method calls are tried in ascending order of their occurrence in the file unless otherwise specified, and if a certain method call involves additional argument(s), we use accessible variables at the faulty statement whose type is compatible with the desired type of each argument to fill in. Second, if there are multiple predicted repair transforms attached to different AST nodes of the buggy statement and the predicted repair transforms are among the 8 repair transforms for which we give the implementation to concretize them into patches, we combinatorially consider the concretization procedure for each involved repair transform. Third, our predicted repair transforms are attached to specific AST nodes, and this can significantly facilitate the synthesis process as the search space will be dramatically reduced. Finally, our synthesizer is by no means complete for the considered 8 repair transforms and more advanced synthesis procedure can generate more diverse and complex patches. Instead, our synthesizer serves to illustrate that our model can precisely predict high-level code transforms attached to AST nodes and these high-level code transforms can effectively guide the concrete code transform process.

\begin{table}
 \caption{The comparison of our proof-of-concept synthesizer with five DL-based APR systems}
  \label{tab:repair}
  \centering
  \small
  \begin{tabular}{|l|c|}
    \hline
    \tabincell{c}{Compared \\ APR System } & \tabincell{c}{Bugs Repaired Only by \\ Our Proof-of-concept Synthesizer } \\
    \hline
    CODIT \cite{CODIT}   & \tabincell{c}{Chart-1, Chart-20, Closure-10, Closure-14, \\ Closure-31, , Closure-70, Closure-123, \\ Math-5,  Math-58, Math-75, Math-85, \\  Mockito-26, Time-19}  \\
        \hline
    CoCoNut \cite{CoCoNuT} & \tabincell{c}{Chart-8, Chart-20, Closure-10, Closure-14, \\ Closure-123, Math-5, Math-70, Math-75, \\ Math-85, Mockito-26}\\
        \hline
    DLFix \cite{DLFix}  & \tabincell{c}{Chart-8, Closure-31,  Closure-123,\\  Mockito-26} \\
        \hline
    CURE \cite{jiang2021cure} & \tabincell{c}{Closure-14, Closure-31,  Closure-123,\\ Math-5,  Math-85, Mockito-26} \\
    \hline
    Recoder \cite{Recoder}  & Closure-123,  Mockito-26 \\
    \hline
\end{tabular}
\end{table}

In line with existing work on automatic program repair \cite{ wen2018context, Elixir, XinASE, YuEmSE}, our proof-of-concept synthesizer considers a patch as correct if it passes the test suite associated with each project. Once the synthesizer gets a plausible patch, it stops the patch synthesis process. We apply the synthesizer solely to concretize the top 1 prediction (of needed AST-level repair transforms) by our model into patches. If the repair transform(s) involved in the top 1 prediction fall(s) into the 8 repair transforms that the synthesizer has considered, the synthesizer proceeds and tries to come up with a patch, otherwise it directly returns "patch synthesis failure" result. As the synthesizer currently does not concretize all the 16 considered repair transforms into patches, this is a reasonable evaluation setting. Note that if we consider concretizing top \emph{n} (\emph{n}>1) predictions and directly ignore the prediction that involves repair transforms for which we currently do not have an implementation to concretize them, over-optimistic results maybe obtained as an actual implementation for these repair transforms may return a plausible patch and prevent the synthesis process for other predictions with lower probability.  Among the 38 Defects4J bugs for which our model gives the perfect prediction (i.e., the 38 bugs shown in column "\#(top-1)" of \autoref{tab:dataset}), the needed repair transforms for 21 bugs fall into the 8 repair transforms that the synthesizer has considered, and the remaining 17 bugs involve the other 8 repair transforms. 

We apply our synthesizer to the 21 bugs, and the synthesizer successfully returns plausible patches for 16 bugs. For the other 5 bugs, more advanced synthesis procedure will be needed. As the test suite is in general incomplete, the generated patch by the synthesizer can just over-fit to the test suite and be plausibly correct \cite{YuEmSE,xinoverfitting,yangoverfitting}. We compare the synthesized plausible patch with the human patch, and find that the synthesized patches for the 16 bugs are same as the human patches. As overfitting is a fundamental issue for program repair techniques, this reflects that by attaching repair transforms to specific AST nodes, our approach has the potential to significantly alleviate the over-fitting issue. 

We compare the results achieved by the proof-of-concept synthesizer with existing deep learning based automatic program repair approaches. To enable fair comparison, we need DL-based APR approaches that have been evaluated under the perfect fault localization, i.e., the buggy statement is known. Note that this setting is the preferred way to evaluate APR approaches according to recent work \cite{testsuiterepair}, as it enables fair assessment of APR techniques independently of the fault localization approach used. Also, we require that the APR approaches are evaluated using \emph{Defects4J v1.2}. Finally, we select 5 state-of-the-art DL-based APR approaches that satisfy the requirement, including CODIT \cite{CODIT}, CoCoNut \cite{CoCoNuT}, DLFix \cite{DLFix}, CURE \cite{jiang2021cure}, and Recoder \cite{Recoder}. According to the original papers, CODIT, CoCoNut, DLFix, and Recoder respectively repair 16, 44, 40, 57, and 67 bugs under the perfect fault localization setting.

\textbf{Results}. We compare the 16 bugs repaired by our proof-of-concept synthesizer with the bugs repaired by these 5 DL-based APR approaches, and find that our synthesizer repairs a significant number of unique bugs as shown in \autoref{tab:repair}). Compared with CODIT, CoCoNut, DLFix, CURE, and Recoder, the proof-of-concept synthesizer repairs 13, 10, 4, 6, and 2 unique bugs respectively. The result implies that on top of the AST-level repair transforms predicted by our model, even a proof-of-concept synthesizer is highly complementary to these best-performing DL-based APR approaches. By using a more advanced and complete synthesis procedure for all the 16 considered repair transforms by our model, we could hypothetically concretize other predicted repair transforms into patches and the patch generation process accordingly can take top \emph{n} (\emph{n}>1) predictions into account before a plausible patch is found. In this way, many more correct patches hypothetically will be generated.

\begin{figure}
\centering
 \includegraphics[width=8cm,height=0.8cm]{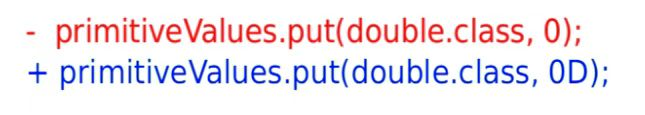}
 \vspace{0.6\baselineskip}   
 \caption{
  The Mockito-26 bug repaired uniquely by our approach. 
}
 \label{fig:repair-example}
\end{figure}

While existing DL-based APR approaches typically use the encoder-decoder architecture and view the patch as a sequence of tokens \cite{DLFix} or edits \cite{Recoder}, our approach for generating patches instead explicitly separates the patch generation process into high-level repair transform prediction and high-level repair transform concretization phases. This is in line with the widely used paradigm adopted by program synthesis techniques \cite{murali2017neural, SQLizer} which separates code synthesis into sketch generation and sketch completion phases. Given that both tokens and edits are low-level changes to some extent, they thus are not quite learnable across projects. For tokens, it is all-too-common that projects can use project-specific identifiers and literals. While edits are higher-level changes than token changes, predicting the exact order of a sequence of edits and the parameters of each edit is also a daunting task. Instead, high-level repair transforms abstract away the project specific details and capture the essential patterns, and are thus much more learn-able across projects. Paired with an effective synthesis strategy, the high-level repair transforms can be concretized into the final patch. Our proof-of-concept synthesizer shows that using our paradigm, even a quite simple synthesis strategy can repair a significant number of unique bugs. As an example, \autoref{fig:repair-example} shows the Mockito-26 bug which is repaired by our synthesizer but not by the 5 considered DL-based APR approaches. To repair the bug, we need to update the constant literal "0" with constant literal "0D". From the perspective of token-level or edit change, this literal update is quite rare in the training data and the DL-based APR approaches do not perform well. Note edit change needs the specific literal "0D" as the parameter. However, from the perspective of high-level repair transform, this kind of constant replacement change (i.e., $\tt{Constant-Rep}$) is much more common and our structured code transform prediction approach does the perfect prediction. Given the predicted $\tt{Constant-Rep}$ transform, our proof-of-concept synthesizer successfully generates the correct patch. 

\textbf{Summary}. Our paradigm for generating patches explicitly splits the patch generation process into high-level repair transform prediction and high-level repair transform concretization phases, and shows clear advantages over existing DL-based APR approaches in case the ground-truth patch involves project-specific low-level (at the level of token or edit) changes that are not learnable across projects. In addition, note that our predicted repair transforms are attached to specific AST nodes, which can guide the patch synthesis in a highly focused manner and the synthesized patches thus suffer less from the overfitting issue.

\subsection{Discussion}
\noindent
\textbf{Evaluation Subjects: Large Diffs over Small Diffs.} {Our evaluation limits the number of root tree edit operations and targets the repair transforms associated with small diffs. 
Note, however, that our approach itself is applicable to ASTs of any code snippet, and we limit the evaluation to small diffs as evaluation on large diffs would pose several problems. First, for large diffs, tree differencing tools like GumTree have accuracy issues, i.e., the produced edit scripts are not accurate enough to reflect real code changes. We refer the readers to \cite{de2018imprecisions} for detailed analysis and concrete examples for the accuracy issues. When the edit script is inaccurate, the repair transforms cannot be correctly attached to AST nodes. Second, a large bug-fixing commit is likely to include changes such as feature additions or refactorings that are irrelevant to bug fixing \cite{just2014defects4j}, and it is notoriously difficult to detect these irrelevant changes \cite{tsantalis2018accurate}. By focusing on small diffs, we significantly increase the chance that the code changes are actually bug-fixing ones. Third, while our defined 16 repair transforms cover the typical repair actions for the five most common repair patterns, some additional repair transforms (e.g., type replacement) may need to be defined in order to fully consider the large bug-fixing diffs, and we consider this as future work.}

\vspace{1mm}
\noindent
\textbf{Performance Diversity over Different Transforms.} 
{We can see from the recorded results that despite our approach achieves overall good performance, its performance varies over different repair transforms. When the overall best performance model is considered (\emph{q} = 0.5) and \emph{top-3} is used as the evaluation metric, while our established CRF model achieves 0.98 accuracy for the $\tt{LogExp-Exp}$ repair transform, the accuracy 0.14 for repair transform  $\tt{Wrap-IFELSE-N}$ is relatively low. We discover that the performance of our established CRF model on different repair transforms is typically in line with that of the \emph{history probability baseline}. In other words, the performance of the established CRF model is relatively good (resp. bad) if that of the \emph{history probability baseline} is not too bad (resp. bad). The implication is that our established CRF model is inclined to assign certain "node-transform" pairs with high probabilities if such "node-transform" pairs are comparably frequent in the training data used to build the model.
The exception is repair transform $\tt{LogExp-Red}$ for which the history baseline achieves accuracy of 0.0 when \emph{top-3} is used as the evaluation metric, but our overall best performance CRF model still has the accuracy of 0.81 using the same evaluation metric. We believe that the underlying reason is that our established feature functions can precisely capture the actual context of the AST nodes that implies this transform. To improve the performance of our model over all repair transforms, one essential way is to further enrich the current feature functions so that they can better capture the relation between AST node context and repair transform. }

\vspace{1mm}
\noindent
\textbf{Learning Time.}
{The CRF model learning process requires calculating the marginal probability many times and our approach uses the exact inference algorithm. The exact inference algorithm is relatively slow and can require exponential time in the worst case for complex graphs, and that is why the training process takes around 41 hours for the full model. To speed up the training process, in particular when we consider large changes for which the input AST will be much larger, we believe that approximate inference algorithms can be used. For CRF model learning, there are two popular classes of approximate inference algorithms \cite{PGM}: Monte Carlo algorithms and variational algorithms. Monte Carlo algorithms attempt to approximately produce a sample from the distribution of interest using a stochastic process. By trying to find a simple approximation that most closely matches the intractable marginals of interest, Variational algorithms instead convert the inference problem into an optimization problem. Studying the applicability and performance of alternative inference
algorithms is an interesting area of future work. }

\vspace{1mm}
\noindent
\textbf{Suitable Application Scenario.} We envision that the usefulness of the predicted code transforms lies more in guiding the automated code evolution in a try-and-error manner. 
The iterative loop compensates for the fact that the predicted code transforms by our model can not always be perfect. For the instantiated particular kind of code transform---repair transform, in line with most program repair techniques, the predicted repair transforms will be embodied into final patches (through synthesis) using exactly the try-and-error manner, called generate-and-validate in the program repair community. That is, the faulty program is iteratively modified with different changes synthesized according to the predicted repair transforms and then validated with a fitness function (typically test suite evaluation) until a plausible patch is generated. Due to the absence of a perfect fitness function \cite{qiicse14}, the full correctness of the plausible patch will have to be established through manual check. Thus, despite the potentially compromised prediction accuracy (i.e., imperfect prediction), the try-and-error workflow can reconcile with the imperfection and effectively generate patches as debugging aids. 

\subsection{Threats to Validity}
\noindent
We leverage the Boa dataset in this study and one potential threat to external validity is whether the results presented in this paper will generalize to other datasets. However, to the best of our knowledge, Boa dataset is one of the largest well established datasets of Java bugs and has been widely used as the experimental subject \cite{megadiff,dyer2014mining,asaduzzaman2016developers}. Another threat to external validity is that we use 16 repair transforms to evaluate the approach and doubts may arise whether the approach will still be effective for other code transforms. Note that our approach itself is not limited to specific types of code transform, it only requires code transforms are defined on AST nodes and we give the framework for defining AST-level code transforms. The 16 considered repair transforms serve as an instantiation of the code transform and they cover the typical repair actions for the five most common repair patterns. A final threat to external validity is that our experimental evaluation uses repair transforms extracted from bug-fixing commits with relatively small diffs, and concerns may be raised about whether the dataset extracted this way is representative enough and whether our approach still has advantages over the two baselines when considering large diffs. Note, however, that our approach itself is applicable to ASTs of any code snippet, and we confine the evaluation to small diffs only for the reasons given in Section 6.6. For the identified small diffs, we do not intentionally select specific ones from which to extract the defined repair transforms. In addition, note that our approach takes code structure into account compared to the two baselines, which is essential for achieving good prediction accuracy regardless of the size of the input AST and the number of repair transforms associated with it. 

A potential threat to internal validity is that our evaluation needs to attach repair transforms to AST nodes and concerns may be raised about the accuracy of this process. To reduce this threat as much as possible, our evaluation focuses on repair transforms associated with bug-fixing commits with relatively small diffs and the manual check suggests that the achieved accuracy is satisfactory, achieving 88\% correctness rate for "multiple-transform" case and at least 91\% correctness rate for "single-transform" case. With regard to possible errors in our implementation and experiments, the whole artifact related with this paper is made available online for scrutiny and extension by other researchers. Besides, we also use 10-fold cross validation to objectively assess the performance of our established model.

\subsection{Future Directions}
\noindent
To begin with, our evaluation in this paper focuses on repair transforms extracted from bug-fixing commits with relatively small diffs for the reasons given in Section 6.6. One important future direction is investigating the performance of the approach on repair transforms associated with large bug-fixing diffs. To achieve this, we need to improve state-of-the-art AST differencing tools and design accurate and efficient refactoring detection algorithms so that we can accurately attach repair transforms to AST nodes according to our definition. In particular, the "\emph{transform number imbalance issue}" is likely to be more serious for large code changes, and we would like to study whether the proposed distribution aware learning can still effectively alleviate this issue and design possible improvements if needed.

Then, our approach lifts AST edits to high-level code transforms and this abstraction brings both reasonable accuracy and scalability for the arguably hard learning task of predicting code changes. Some high-level code transforms will involve selecting certain code elements to operationalize them and another future direction is designing customized effective synthesis algorithm to finish this process. Our proof-of-concept synthesizer concretizes 8 repair transforms to get the final patches, it is worthwhile to explore more advanced and complete synthesis procedures that account for all the 16 considered repair transforms by our model.
There are abundant effective code snippet synthesis algorithms \cite{gvero2015synthesizing, perelman2012type,galenson2014codehint,codecompletionPLDI,raghothaman2016swim} that we can potentially build upon. 

Besides, our evaluation in this paper assumes that we know the faulty code snippet in advance. Thus, it would be interesting to investigate the performance of our approach when it is integrated with fault localization techniques \cite{yuicse,yuist,YUGUI}. In this way, we can more objectively evaluate its effectiveness in industry and overcome potential limitations.

In addition, the realization of our approach currently uses repair transform for fixing bugs as instantiation of code transform, it would be worthwhile to investigate the effectiveness of our approach for other kinds of transforms such as code refactoring and feature improvement.

Furthermore, another important future direction is recognizing additional factors other than program syntax and semantics that can influence code evolution, and integrate them into the probabilistic model. For instance, modern software development typically features the use of integrated development environments (IDEs) and there are in general a broad range of tools within IDEs for helping developers with the code transform task \cite{murphy2011we,ni2021recode,7884626}. It would be valuable to explore the integration of these programming environment factors into our established model. To explore such a direction, one important obstacle we need to overcome is acquiring a great deal of code evolution data associated with programming environment information. 

Finally, our evaluation in Section 6.3 shows that the use of both observation-based and indicator-based feature functions are indispensable for obtaining good prediction performance. For observation-based feature functions, we respectively consider various type related, usage related, and syntax related program element characteristics. For indicator-based feature functions, we respectively consider structural features for AST vertices, edges, and triangles. It would be worthwhile to detailly investigate the specific role each of these code features plays in obtaining the good prediction performance, and a feasible way to achieve this is employing the Recursive Feature Elimination (RFE) method \cite{huang2018feature, darst2018using}. 

\section{Related work}
\noindent
In this section, we review some work closely related with this paper. 

\subsection{Big Code} 
\noindent
{"Big code" approaches \cite{property-predict-popl} for programming refer to data-driven approaches that typically learn form the large amount of freely accessible code, in particular in on-line repository hosting services such as GitHub and BitBucket. The research interest on "Big code" approaches has been rising rapidly in recent years, and we can already see a great many successful applications. Some example applications include defect prediction \cite{4027145,menzies2010defect}, code completion \cite{Bruch2009,codecompletionPLDI}, code synthesis \cite{gvero2015synthesizing}, name-based bug detection \cite{namebug}, deobfuscation \cite{binarypredict}, inferring cryptographic API rules \cite{paletov2018inferring}, effort estimation \cite{8255666}, generation of meaningful assert statements for unit tests\cite{Watson2020OnLM}, detection of argument selection bug \cite{10.1145/3133928}, infer of loop invariants \cite{10.5555/3327757.3327873}, type inference for dynamically typed languages \cite{10.1145/3385412.3385997}, discovery of aliasing specifications of APIs \cite{eberhardt2019pldi}, acquisition of strategy for adapting program analysis \cite{oh2015learning}, fixing MOOC programs \cite{pu2016sk_p}, 
sketch implementations \cite{ICLRSynthesis} or learn the relation between code features and sketch implementations \cite{murali2017neural}, and fuzzing smart contracts \cite{10.1145/3319535.3363230}, etc. We refer the readers to \cite{allamanis2018survey} for a recent overview of machine learning on code.}

In this paper, we target a problem that has not been studied before: the prediction of code transforms at the level of AST nodes. In the next two sections, we respectively review the closely related areas about "code transform learning" and "program representation learning". 

\subsection{Code Transform Learning} 
\noindent
During the life-cycle of a software, its source code will go through a lot of transformations, for reasons such as bug fixing, code refactoring, and feature improvement. To aid in the code transform process, a lot of learning-based techniques have been developed. 

Meng et al. \cite{MengPLDI} propose an approach which learns from a similar but not identical edit instance to create a context-aware edit script. They later further extend the approach to learn from multiple edit instances, enabling automatic identification of the target edit location and generation of more accurate edit script \cite{meng2013lase}. Rolim et al. \cite{rolim2017learning} use a programming-by-Example (PBE) technique to search for code transform that is consistent with all example transform instances. Long et al. \cite{Gensis} infer code transforms from bug-fixing commits that have fixed the same type of exception bugs. For automating API migration, by learning from client differences in API usage from multiple examples, Andersen et al. \cite{andersen2010generic} create an edit script for the correct API usage and then update the incorrect API use accordingly. Koyuncu et al. \cite{koyuncu2020fixminer} make use of an iterative clustering strategy to mine fix patterns from atomic changes within patches. To make programming by demonstration/examples (PBD/PBE) systems more widely adopted in practice, Miltner et al. \cite{miltner2019fly} present a modeless system that can synthesize context-specific repetitive edits on the fly. Bader et al. \cite{bader2019getafix} propose a hierarchical clustering algorithm that summarizes past fix patterns into a general-to-specific hierarchy and then a context-aware ranking technique is used to select the most appropriate fix for a specific bug. The approach targets six bug categories, such as null dereferences and incorrect API calls. When using multiple edit instances, these approaches require either that all the example edits are similar enough or that the example edits are clustered (manually or automatically) using a certain similarity criterion first. Essentially, these approaches try to extract useful edit information from a few highly similar (according to certain definition of similarity) edit instances. However, it can be difficult for few highly similar examples to contain all the information needed for making prediction. In contrast, our presented approach is able to extract and assembly information from diverse (i.e., not that similar) example edits and collectively builds a probabilistic model that can make accurate prediction for general categories of code transform.

There are few recent works that aim to establish probabilistic models of code transforms. Given a partially finished edit, Brody et al. \cite{brody2020structural} aim to predict the remaining parts of the edit. The technique tries to represent the four basic AST edit operations (\text{UPD}, \text{ADD}, \text{DEL}, \text{MOV}) using two nodes from the original AST, and then uses the AST path between the two nodes to represent an edit operation. Unlike our approach which in principle is amenable to all kinds of edit operations, an issue with their technique is that a significant portion of real-life edit operations can not be represented using their representation. Yin et al. \cite{yin2018learning} propose to use a neural architecture to embed source code edits in a high-dimensional embedding space. One key difference with our work is that their embedding is not actionable. Conversely, our approach abstracts basic tree edit operations into code transforms on AST nodes that are fully actionable and explainable. Besides, unlike our work, the work of Yin et al. \cite{yin2018learning} typically requires that the edit to be predicted is exactly included in the context. Tufano et al. \cite{tufano2019empirical} mine bug fixes from GitHub projects and use the associated AST edit actions to train an Encoder-Decoder model capable of translating buggy code into correct code. However, one limitation is that the translation model can only learn fixes that involve re-arranging keywords, identifiers, and literals already available in the context of the buggy code. Mesbah et al. \cite{deepdelta} train a neural network to learn simplified edit script. Unlike our work which targets the wide variety of code errors and the predicted code transforms are associated to specific AST nodes, the work of Mesbah et al. \cite{deepdelta} focuses on certain compilation errors and the learned edit script does not contain edit position information. More importantly, compared to all of these works, only our approach abstracts the AST edits into high-level code transforms and this abstraction brings both accuracy and scalability for this hard learning task.

\subsection{Program Representation Learning}
\noindent
Within the big code literature, some works specifically target the representation problem. In particular, inspired by the word2vec~\cite{mikolov2013efficient} milestone, which embeds words into a numerical space where similar meaning words are in close proximity, many research efforts have been devoted to neural program representations in recent years.

Alon et al. \cite{alon2019code2vec} use a path-based attention model to learn continuous distributed vector representations for code snippet (i.e., code embeddings), which then can further be used to predict semantic properties of the snippet. It is yet not clear how the learned representation can be used for predicting the newly defined AST-level code transforms in this paper. DeFreez et al. \cite{defreez2018path} use paths in the control flow graph to embed functions, aiming to identify function synonyms. Allamanis et al. \cite{programgraphlearning} explore a graph-based representation of both the syntactic and semantic structure of code, which then is given as input to a graph neural-network. Henkel et al. \cite{henkel2018code} represent programs with abstractions of traces obtained from symbolic execution. Hellendoorn et al. \cite{hellendoorn2017deep} argue that source code has special properties and improve existing language modeling techniques by incorporating handle of particular aspects of source code, including for example frequent changes and deeply nested scopes. Ideally, part of the representation is automatically inferred for these works, which is called "learned semantic feature" by Wang et al. \cite{astlearndefectprediction}. Compared with neural embedding of code snippet, our approach abstracts code edits into code transforms on AST nodes that are fully actionable and explainable. Our representation of programs is a combination of abstract syntax trees (AST) and rich carefully engineered and powerful features, which not only facilitates the learning process but also enables the definition of explainable AST-level code transform. 

\subsection{Data-driven Program Repair}
\noindent 
The past two decades have witnessed an ever-growing research interest on program repair, which aims to automatically correct bugs. Two most well-known categories of program repair techniques are generate-and-validate techniques \cite{le2012genprog} and synthesis-based techniques \cite{nguyen2013semfix}. We refer the readers to \cite{Monperrus2015} for an overview of program repair.

In particular, some works in the program repair literature \cite{Monperrus2015} explore data-driven program repair. To investigate the effectiveness and promise of program repair in fixing real bugs, Zhong et al. \cite{zhong2015empirical} empirically study more than 9,000 real-world bug fixes and come up with four insights on fault localization \cite{yuicse,yuist} and faulty code fix. Le et al. \cite{Le2016HDRepair} use the commit history to select the most likely mutation-based patch. Long et al. \cite{Prophet} propose an  approach to select the patch that resembles the most to past human patches, aiming to get patches that suffer less from overfitting \cite{YuEmSE}. Wang et al. \cite{feedbackgeneration} use data-driven program repair for automated feedback generation. Jiang et al. \cite{jiang2018shaping} combine the search space from similar code with search space obtained by abundant existing patches to better guide program repair. Wen et al. \cite{wen2018context} use three AST mutation operators (Replacement, Insertion, and Deletion) to make the search space include potentially more correct patches, and use context information of fixing ingredients obtained from real bug fixes in open source projects to explore the search space. 
Chakraborty et al. \cite{CODIT} propose CODIT, which learns code change patterns by employing a tree-based neural machine translation model to encode source code changes.
Lutellier et al. \cite{CoCoNuT} make use of ensemble learning on the combination of
convolutional neural networks (CNNs) and 
neural machine translation (NMT) model to generate patches token by token. 
Li et al. \cite{DLFix} introduce a two-tier DL model named DLFix, whose first layer aims to learn the contexts of bug fixes and second layer tries to learn the bug-fixing patch by using the result from first layer as an additional weighting input.
Jiang et al. \cite{jiang2021cure} propose CURE, which makes use of three strategies to to improve neural machine translation based APR techniques, including pre-trained programming language model, code-aware search, and sub-word tokenization.
Zhu et al. \cite{Recoder} propose Recoder, which uses a syntax-guided manner to generate edits
and features provider/decider architecture and placeholder generation.
In our work, we represent code with full ASTs and carefully engineered code features, and combine the representation with structured prediction to predict AST-level repair transforms. The predicted high-level repair transforms are then concretized into the final patches on top of effective code snippet synthesis algorithms. To our knowledge, this radically departs from the existing related work in data-driven program repair.

\subsection{Conditional Random Fields for Programming} 
\noindent
{Conditional random fields (CRFs) belong to the powerful probabilistic graphical models (PGM) and have been successfully used in a wide variety of applications, including information retrieval \cite{CRFapplication2}, natural language processing \cite{CRFnlp}, bioinformatics \cite{CRFbiomedical}, and computer vision \cite{CRFapplication1}. Recent years have also witnessed a growing interest of using CRFs for programming tasks, and we finally review representative works in this direction.}

{By modelling programs with CRFs, Raychev et al. \cite{property-predict-popl} achieve good results in predictions of identifier names and variable types for JavaScript code. To establish relationships between program elements, they use certain grammar rules peculiar to JavaScript. He et al. \cite{binarypredict} build CRFs for assembly code and then use the established CRF model to predict properties of meaningful elements in unseen stripped binaries, including symbol names, types and locations. Similarly, Bichsel et al .\cite{ccs16} target Android applications and employ CRFs to achieve prediction of identifier names renamed by layout obfuscation. Given the increased availability of software engineering social content (e.g, Stack Overflow Q\&A discussions), Ye et al. \cite{Yesaner} adapt CRFs for analyzing software engineering texts such that a broad category of software entities for a wide range of popular programming languages, platforms, and libraries can be recognized. Differently, our way of establishing the CRF model for code is novel and our use of CRFs for predicting code transforms is fundamentally original.}

\section{Conclusion}
\noindent
A computer program evolves under a sequence of code transforms, for reasons such as bug fixing and code refactoring. These code transform activities represent a significant amount of the cost of a system, calling for tools to aid them. We present in this paper the first approach for structurally predicting code transforms at the level of AST nodes. Our approach leverages conditional random fields (CRFs), and first learns offline a probabilistic model that captures how certain code transforms are applied to certain AST nodes from existing data, and then uses the learned model to predict transforms for arbitrary new, unseen code snippet. Our approach is generic for different kinds of code transforms and languages, and involves novel representation of both programs and code transforms. {In particular, we introduce the formal framework for defining AST-level code transforms and we demonstrate how the CRF model can be accordingly designed, learned, and used for prediction.} We present in this paper an instantiation in the context of repair transform prediction for Java programs. Our instantiation contains a set of carefully designed code features, and deals with the training data imbalance and transform constraint issues that arise for repair transform prediction problems. A large-scale experimental evaluation has shown the effectiveness of our approach. The result also shows that our model outperforms two baselines based on history probability and neural machine translation (NMT), suggesting the importance of considering code structure in achieving good prediction accuracy for AST-level code transform. In addition, a proof-of-concept synthesizer is implemented to concretize some repair transforms to get the final patches. The evaluation of the synthesizer on the Defects4j benchmark confirms the usefulness of the predicted AST-level repair transforms in producing high-quality patches.

\section*{Acknowledgments}
\noindent
We are grateful to the anonymous reviewers for their insightful comments. This work was partially supported by National Natural Science Foundation of China (Grant No. 62102233), Shandong Province Overseas Outstanding Youth Fund (Grant No. 2022HWYQ-043), Qilu Young Scholar Program of Shandong University, and Wallenberg Artificial Intelligence, the Wallenberg Autonomous Systems and Software Program (WASP) funded by Knut and Alice Wallenberg Foundation. Some experiments were performed on resources provided by the Swedish National Infrastructure for Computing.


\end{document}